\documentclass[pra,twocolumn,superscriptaddress,floatfix,amsmath,amsfonts,amssymb,longbibliography]{revtex4-1}%
\usepackage{amsmath,amsfonts,amssymb,color}
\usepackage{amsthm}
\usepackage{leftidx}
\usepackage{graphicx}
\usepackage{subfigure}
\def\GJ{\textcolor{black}}
\def\CH{\textcolor{black}}
\def\TT{\textcolor{black}}
\usepackage{xcolor}
\usepackage{dcolumn}
\usepackage{bm}
\usepackage{epstopdf}
\usepackage{epsfig}
\usepackage{physics}
\begin{document}

	\title{Nonlinearity induced topological physics in momentum space and real space}
	\author{Thomas Tuloup}
	\affiliation{%
		Department of Physics, National University of Singapore, Singapore 117543
	}
	\author{Raditya Weda Bomantara}
	\email{Raditya.Bomantara@sydney.edu.au}
	\affiliation{Centre for Engineered Quantum Systems, School of Physics, University of Sydney, Sydney, New South Wales 2006, Australia}
\author{Ching Hua Lee}
\affiliation{%
	Department of Physics, National University of Singapore, Singapore 117543
}
	\author{Jiangbin Gong}%
	\email{phygj@nus.edu.sg}
	\affiliation{%
		Department of Physics, National University of Singapore, Singapore 117543
	}
	\date{\today}
	
	%%%%%%%%%%%%%%%%%%%% ABSTRACT %%%%%%%%%%%%%%%%%%%%%%%%
	%\begin{linenumbers}
	
	\vspace{2cm}
	
\begin{abstract}
Nonlinearity induced topological properties in nonlinear lattice systems are studied in both momentum space and real space.  Experimentally realizable through the Kerr effect on photonic waveguide systems, our working model depicts on-site nonlinearity added to the Su-Schrieffer-Heeger  (SSH) model plus a chiral-symmetry breaking term.    Under the periodic-boundary condition, two of the nonlinear energy bands approach the energy bands of the chiral-symmetric SSH model as nonlinearity strength increases.  Further, we account for a correction to the Zak phase and obtain a general expression for nonlinear Zak phases.  For sufficiently strong nonlinearity,  the sum of all nonlinear Zak phases (not the sum of all conventional Zak phases) is found to be quantized.   In real space,  it is discovered that  there is a strong interplay between nonlinear solitons and the topologically protected edge states of the associated chiral-symmetric linear system.  Nonlinearity can recover the degeneracy between two edge soliton states, albeit a chiral-symmetry breaking term.  We also reveal the topological origin of in-gap solitons even when the associated linear system is in the topological trivial regime.   These momentum-space and real-space results have clearly demonstrated new topological features induced by nonlinearity, indicating that  topological physics in nonlinear lattice systems is far richer than previously thought.
\end{abstract}
%\pacs{03.65.Vf, 05.60.Gg, 05.30.Rt, 73.20.At}
	
\maketitle
	
\section{Introduction}
	
\GJ{Topological phases of electronic systems and metamaterials have recently been a subject of tremendous theoretical and experimental interests~\cite{Thouless1982,Thouless1983,Kane_and_Mele2005, ExperimentalRealization2007,chen2009experimental,chang2013experimental,Khanikaev2013,Gao2013,lv2015experimental,xu2015discovery,yang2015weyl,Chiu2016,he2017chiral,gao2018experimental,imhof2018topolectrical,schindler2018higher,xue2019acoustic,hofmann2020reciprocal}. Though studies in non-interacting systems have been extremely fruitful, our new knowledge of topological physics in interacting systems~\cite{lopez1991fractional,sorensen2005fractional,bernevig2009clustering,Interaction_Topological_Classification_Fermion,seidel2010s,neupert2011fractional,Classification_ElectroTI_3d,lee2015geometric,Interaction_Classification_CrystalTIandSuperConductor}
is important for both academic pursuit and future innovations.  Indeed, even excluding the Coulomb interaction in solid-state systems, controllable interaction is also  ubiquitous in a variety of platforms exploited to synthesize artificial topological matter, such as the Hubbard interaction in cold atom systems as well as the Kerr-effect in optical and acoustic setups.    %A mean-field approach can also still be useful, usually giving rise to nonlinear Schrodinger equations \cite{ExampleMeanFieldBEC,ChiralityGapSoliton}, e.g., the Gross-Pitaeviskii equation in treating Bose-Einstein condensates in the mean-field limit in cold-atom systems \cite{Review_BEC_2003,ReviewUltracoldAtoms}.  We then expect the theoretical predictions from this paper to be relevant in photonic systems and cold-atom systems in optical lattices.
However, an apparent and inherent difficulty in treating interacting systems lies in the computational complexity of many-body systems, and as such examining topological effects in interacting systems can be theoretically challenging and computationally costly, often requiring the use of advanced many-body techniques and/or sophisticated numerical methods.} % potentially obscuring intuitive insights.

\GJ{As already suggested by the title of this paper, here we adopt a reserved mean-field approach such that  many-body interacting problems are reduced to single-particle nonlinear ones, whose behavior is then governed by certain nonlinear Schr\"{o}dinger equations.  Such treatment is well known, e.g., in handling the mean-field behavior of Bose-Einstein condensates of Bosonic cold atoms~\cite{ExampleMeanFieldBEC,ChiralityGapSoliton,Review_BEC_2003,ReviewUltracoldAtoms} where the nonlinear Schr\"odinger equation is known as the Gross-Pitaevskii (GP) equation~\cite{GrossBEC_GP,PitaevskiiBEC_GP}.    
Studies of nonlinear problems of this type have been extensively made over recent years, in cold-atom systems~\cite{DarkSolitonsBEC,GenerateSolitonsPhaseEngiBEC,MatterWaveSolitonTrains} and photonic metamaterials with optical Kerr effects~\cite{NonlinearTopologicalPhotonics,TopologicalNatureNLOpticalEffectsSolids,SelfLocStatesPhotnicTI,PhotonicGraphene,KerrEffectNonlinearTopologicalPhotonics,EdgeSoltionsNonlinearPhotonicsTI,TopoTransitioninSSHthroughNLKerr,ExampleNonlinearProtectionPossiblePhaseTransition}. 
Important physical features unique in nonlinear systems have been found, such as the emergence of looped structures in their nonlinear energy bands~\cite{NL_LZ_Tunneling,Review_BEC_2003,ReviewUltracoldAtoms, Zhang2008,Lim2020} and the existence of strongly localized solitons~\cite{ChiralityGapSoliton}.}

\GJ{Given that nonlinear lattice systems are closely related to available experimental platforms,  nonlinear lattice systems are hence not just mean-field approximations of certain complex many-body systems; they offer important opportunities to explore novel physics in their own right. Indeed, recent years have witnessed a shifted interest towards the topological aspects of nonlinear lattice systems~\cite{NLandTopology}, with early investigations mostly made through the dynamics of edge states therein~\cite{EdgeSoltionsNonlinearPhotonicsTI,TopoTransitioninSSHthroughNLKerr,NL2ndOrderTI,ExampleNonlinearProtectionPossiblePhaseTransition}. One exception is a study by two of the present authors and others,  where a topological invariant associated with the bulk~\cite{NonlinearDiracCones} was used to characterize a novel type of Dirac cones induced by nonlinearity. 
Motivated by these recent developments, here we aim to advance current understanding of nonlinear topological systems by looking into one-dimensional (1D) nonlinear lattices, with both momentum-space and real-space treatments.}

\GJ{Specifically, we consider a nonlinear SSH model with on-site nonlinearity and a chiral-symmetry breaking term. This system can be realized via photonic systems assembled by waveguides with a Kerr medium. Our key findings are as follows. } 

\GJ{Firstly, the system is investigated in the momentum space under periodic boundary condition (PBC). Instead of using the conventional Zak phase associated with an energy band to seek possibly new topological features due to nonlinearity,  we advocate to use the so-called nonlinear Zak phase, which can account for an additional geometric contribution arising from the adiabatic following of a nonlinear system. Remarkably, it is found that the nonlinear Zak phases, though not quantized individually, can still yield a quantized value when summed over all the nonlinear energy bands, for sufficiently large nonlinear strength. This result unexpectedly reveals a quantized quantity in nonlinear lattice systems, suggesting a possible topological characterization unique to nonlinear systems.}  

\GJ{Secondly, the system is examined in real space under open boundary condition (OBC). As the strength of nonlinearity increases, we observe that the original linear energy spectrum, which comprises delocalized bulk states and localized edge states, breaks down into soliton states, localized either in the bulk or at the lattice edges. Interestingly, the emergence of these solitons can be explained by the idea of self-consistent, nonlinearity induced edges inside the bulk, leading to fascinating examples featuring the interplay between nonlinearity and topology. Consequently, the behaviour of such nonlinear systems can be now largely understood in terms of the topological properties of the associated chiral-symmetric linear system. Solitons existing in the energy gap are particularly engaging, as they exist in a regime where nonlinearity is strong enough to have an effect, but does not completely overrun the features of the original linear model. In this case, the induced edge in the bulk is found to accommodate edge states on each of the two sides of the soliton, in the same fashion as in the linear chiral-symmetric SSH model. Related to this key insight, we observe and explain how nonlinearity with moderate strength leads to the recovery of edge-state degeneracy despite chiral-symmetry breaking}. %an extremely interesting property of the nonlinear system under OBC, which is the effective recovery of the degeneracy of the edge states, cancelling the effect of chiral symmetry breaking.

\GJ{This paper is organized as follows. In Sec.~\ref{section:Computational tools}, we introduce our major theoretical and computational tools respectively for momentum-space and real-space treatments.  Of particular interest is the introduction of a rather general theory of nonlinear Zak phase.
 In Sec.~\ref{section:NLSSHmodel} we describe our working model as a nonlinear SSH model with chiral symmetry breaking. The main results are presented in Sec.~\ref{section:PBC} and Sec.~\ref{section:OBC} from both momentum-space and real-space perspectives. Major results from the momentum-space treatment include nonlinear band structure, behavior of nonlinear and conventional Zak phases, the recovery of quantized Zak phases over a summation over all bands for sufficiently strong nonlinearity, and an analysis of the dynamical stability of the nonlinear energy bands. Major results from the real-space treatment include analysis of the OBC spectrum, localization properties of soliton solutions, the relevance of topological edge states in the linear limit to the interesting profile of in-gap solitons, and a recovery of degeneracy between edge soliton states at the two ends of the nonlinear lattice despite chiral-symmetry breaking. Section \ref{section:Conclusion} summarizes the main findings of this paper, along with suggestions for possible further studies.}

\section{Theoretical and computational tools}
\label{section:Computational tools}

\subsection{Theory of nonlinear Zak phase}
\label{section:Nonlinear Zak phase}

We begin by introducing a general theoretical tool to treat nonlinear lattice systems in the momentum space under PBC. Consider first topological properties of 1D chiral-symmetric linear systems, which can be well characterized by the Zak phase~\cite{ExplainZakPhase} of their bulk energy bands. Here, the Zak phase is defined as the Berry phase associated with the adiabatic evolution of a bulk energy eigenstate as the quasimomentum $k$ is scanned over the Brillouin zone $k = 0 \rightarrow 2 \pi$. In particular, for a two-band system described by the general Hamiltonian %it is equal to half the solid angle spanned by an energy eigenstate on the Bloch sphere as $k= 0 \longrightarrow 2 \pi$, thus giving valuable geometric insight into its topology. In particular, given that any two-level system's Hamiltonian can be written as
\begin{equation}
\label{eqn:H_2level}
\begin{aligned}
H &= \begin{pmatrix} \cos{\theta} & \sin{\theta}e^{-i\phi} \\\sin{\theta}e^{i\phi} & -\cos{\theta} \end{pmatrix} \;,
\end{aligned}
\end{equation}
where $\theta$ and $\phi$ are the angles used to represent the eigenstates on the Bloch sphere (which generally depend on $k$), its Zak phases can be immediately obtained as $\gamma_\pm=\pm\frac{\Omega}{2}$, where $\Omega$ is the solid angle covered by one of its eigenstates as $k$ varies from $0$ to $2\pi$, and $\pm$ labels its two eigenstates. In the presence of chiral symmetry such that $\sigma_z H \sigma_z = -H$, $\cos(\theta)$ is necessarily $0$, and the eigenstates are then bound to evolve in the $x,y$-plane, i.e., on the equator of the Bloch sphere, which yield a quantized Zak phase equal to an integer $n$ multiple of $\pi$, where $n$ represents the number of times the azimuthal angle $\phi$ winds around the origin as $k$ varies from $0$ to $2\pi$. 

The direct connection between Zak phase and winding number above, which highlights the topological nature of such a system, relies heavily on the presence of the chiral symmetry. Perturbations of the form $v\sigma_z$ suffice to break such a symmetry and subsequently the quantization of the Zak phase. In this case, the Zak phase can take any value in $[0,2\pi)$ and thus no longer describes a topological quantity.  \GJ{As shown later,  an intriguing interplay between nonlinearity and chiral-symmetry breaking can be examined via the recovery of almost quantization or even \TT{exact (up to a numerical error of $10^{-7}$)} quantization of a different geometric phase accounting for contributions from nonlinearity.} 

To generalize the definition of Zak phase in 1D nonlinear two-band systems, we first recall that as the quasimomentum $k$ adiabatically runs over one cycle in the Brillouin zone, the total phase acquired by an eigenstate is the sum of two terms,  the dynamical phase and the geometric phase. The dynamical phase is identified as the term arising due to the contribution from the state's time evolution which thus depends on the total time taken to complete the adiabatic cycle, whereas the geometric phase is independent of such a total time and solely depends on the closed path in parameter space (e.g. $\theta$ and $\phi$). \GJ{Interestingly, such a natural division between the geometric phase and the dynamical phase becomes problematic in nonlinear systems. In particular, though the conventional Zak phase in a two-band system (determined by the solid angle traced out by the adiabatic nonlinear eigenstates) still contributes to the geometric phase as in linear systems,  the dynamical phase in nonlinear systems also accumulates a geometrical phase contribution~\cite{NonlinearDiracCones,NonlinearZakPhase}. For this reason below we explicitly develop a theory of nonlinear Zak phase}.

Consider a nonlinear time-dependent Schr\"{o}dinger equation
\begin{equation}
\label{eqn:NLSchrodinger}
\mathrm{i}\hbar \frac{\partial}{\partial t}\Psi = H(\Sigma) \Psi\;,
\end{equation}
where we have defined a nonlinear (state-dependent) ``Hamiltonian"     
\begin{equation}
\label{eqn:GPHamiltonian}
H(\Sigma) = h_x \sigma_x + h_y \sigma_y + h(\Sigma) \sigma_z,
\end{equation}
$\Psi=\left(\Psi_1,\Psi_2\right)^T$, $\Sigma=|\Psi_2|^2-|\Psi_1|^2$, $h_x$ and $h_y$ are assumed to be state independent for simplicity, whereas $h$ can be any function of $\Sigma$. By writing the solutions to Eq.~(\ref{eqn:GPHamiltonian}) as $\Psi(t)=e^{i f(t)} \Phi(t)$ with $f(t)$ being the total phase resulting from time-evolution, we identify $\Phi(t)$ as an element of a projective Hilbert space. By multiplying Eq.~(\ref{eqn:NLSchrodinger}) from the left with $\Psi^\dagger$ and simplifying it, we obtain (summation of repeated indices is implied) %Associated to Eq.(\ref{eqn:GPHamiltonian}), we obtain a relation about the total phase (implying summation of repeated indices)
\begin{equation}
\label{eqn:dfoverdt}
\frac{df}{dt} = i\Phi_a^{*}\frac{d\Phi_a}{dt} - \Phi_a^{*}H_{ab}\Phi_b \;.
\end{equation}
Upon integrating the above with respect to time, the first term on the right hand side is what we normally identify as the Aharonov-Anandan (AA) phase~\cite{AharonovAnandan}, which is usually associated with the (nonadiabatic) geometric phase in linear systems. In nonlinear systems, however, the second term may contain additional geometric contribution. In the adiabatic limit, this in turn modifies the general form of the system's Zak phase. 

By perturbatively expanding both $f$ and $\Psi_a$ under an adiabatic parameter $\epsilon$ as 
\begin{equation}
\label{eqn:AdiabaticExpansion}
\begin{aligned}
\frac{df}{dt} &= \alpha_0 + \alpha_1 \epsilon + \alpha_2 \epsilon^2 + ... \\
\Phi_a &= \Phi_a^{(0)} + \epsilon \Phi_a^{(1)} + \epsilon^2 \Phi_a^{(2)} + ...    \;,
\end{aligned}
\end{equation}
we choose a state initially in a stationary state $\Phi^{(0)}=\Phi_E$ such that $H \Phi_E=E \Phi_E$. During an adiabatic process, the trajectory of the state $\Psi^{(0)}$ gives rise to the conventional Zak phase defined in linear systems. In linear systems, this is also the only geometric contribution, since variations in the dynamical phase contribution of Eq.~(\ref{eqn:dfoverdt}) will only yield terms that are at least of order $\epsilon^2$, which vanish in the adiabatic limit. On the other hand, since $H$ is state dependent in nonlinear systems, its variation induced by the time-evolution of the state yields a term in the dynamical phase contribution of Eq.~(\ref{eqn:dfoverdt}) that is of first order in $\epsilon$, thus giving rise to another geometric contribution. In particular, by substituting Eq.~(\ref{eqn:AdiabaticExpansion}) into Eq.~(\ref{eqn:dfoverdt}), then evaluating zeroth and first order terms in $\epsilon$, we obtain %However, due to the nonlinear dynamics, this conventional Zak phase never appears alone. Indeed, the fact that a change in the state induces a change in the Hamiltonian signifies that no matter how slow the adiabatic process is, there will be perturbation in the linear order of $\epsilon$ in the dynamical phase, a second geometric contribution. In the linear case, there is no geometric contribution from the dynamical phase for a slow enough adiabatic process, as all perturbations are at least of order $\epsilon^2$, but we obtain here in the adiabatic limit 
\begin{equation}
\label{eqn:GeneralAdiabaticLimit}
\begin{aligned}
\alpha_0 &= -E,\\
\epsilon \alpha_1 &= \underbrace{i \Phi_a^{(0)*} \frac{d\Phi_a^{(0)}}{dt}}_{\substack{\textrm{Original Berry} \\ \textrm{connection}}} \underbrace{- \epsilon \Phi_a^{(0)*} H_{ab}^{(1)} \Phi_a^{(0)}}_{\substack{\textrm{Geometric contribution} \\ \textrm{from dynamical phase}}},
\end{aligned}
\end{equation}
where $H^{(1)}=\left. \frac{dh}{d\Sigma} \right\rvert_{\Sigma=\Sigma^{(0)}} \left.\frac{d\Sigma}{d\epsilon} \right\rvert_{\epsilon=0} \sigma_z$, and the absence of $\epsilon$ in the first term on the right hand side of $\epsilon\alpha_1$ in  Eq.~(\ref{eqn:GeneralAdiabaticLimit}) is due to the fact that $\frac{d\Phi_a^{(0)}}{dt}\propto \epsilon$ in the adiabatic limit. For the two-level nonlinear Hamiltonian described in Eq.~(\ref{eqn:GPHamiltonian}), this means 
%\begin{equation}
%\label{PerturbativeExpansionHamiltonian}
%H = H^{(0)} + \epsilon H^{(1)} + \epsilon^2  H^{(2)}... 
%\end{equation}
\begin{equation}
\label{eqn:PerturbativeAlpha}
\begin{aligned}
\alpha_0 &= -E,\\
\epsilon \alpha_1 &= i \Phi_a^{(0)*} \frac{d\Phi_a^{(0)}}{dt} - 4 \epsilon \left. \frac{dh}{d\Sigma} \right\rvert_{\Sigma^{(0)}} \Sigma^{(0)} \operatorname{Re}\left(\Phi_1^{(0)*} \Phi_1^{(1)}\right),
\end{aligned}
\end{equation}
where $\Sigma^{(0)}=\left|\Phi_2^{(0)}\right|^2-\left|\Phi_1^{(0)}\right|^2$ and normalization condition $\Re(\Phi_a^{(1)*}\Phi_a^{(0)})=0$ has been employed in the above. The stationary state $\Phi_E$ can further be written without loss of generality \footnote{\label{ftn:Choice2levelGauge}Although choosing $\ket{\Phi^{(0)}} = \begin{pmatrix} \cos{\frac{\theta}{2}} \\ \sin{\frac{\theta}{2}} e^{i\phi} \end{pmatrix}$ seems to assume that we are restraining ourselves to stationary states with positive energy $E>0$, we can simply obtain all the equivalent results for $E<0$ with $\ket{\Psi^{(0)}} = \begin{pmatrix} \sin{\frac{\theta}{2}} \\ -\cos{\frac{\theta}{2}} e^{i\phi} \end{pmatrix}$ by changing $\theta \rightarrow \theta' = \theta + \pi$ in all our expressions} in the form
\begin{equation}
\label{eqn:StationaryPsiE}
\Phi_E=\begin{pmatrix} \cos{\frac{\theta}{2}} \\ \sin{\frac{\theta}{2}}e^{i\phi} \end{pmatrix}.
\end{equation}
After substituting it into Eq.~(\ref{eqn:PerturbativeAlpha}), and doing some algebra detailed in Appendix~\ref{app}, we obtain the first order term of $\frac{df}{dt}$ as a Berry connection modified by a kernel $\mathcal{K}$ deforming a familiar intergral.
\begin{equation}
\label{eqn:CorrectionNLBP}
\begin{aligned}
\epsilon \alpha_1 &= \mathrm{i} \mathcal{K}\Phi_a^{(0)*}\frac{d\Phi_a^{(0)}}{dt} \;, \\
\mathcal{K} &= \left( 1 + \frac{\left. \frac{dh}{d\Sigma} \right\rvert_{\Sigma^{(0)}} \cos{\theta} (1+\cos{\theta})}{E +  \left. \frac{dh}{d\Sigma} \right\rvert_{\Sigma^{(0)}} \sin^2{\theta}} \right) \;.
\end{aligned}
\end{equation}
Consequently, the nonlinear Zak phase for any 1D two-band systems with diagonal nonlinearity $h(\Sigma)$ is given by
\begin{equation}
\label{eqn:ExpressionNLBP}
\gamma_{NL} = \int_{0}^{2\pi} \mathrm{i}\mathcal{K}(k)\Phi^{(0)}_a(k)^* \frac{d \Phi_a^{(0)}(k)}{dk} dk \;,
\end{equation}
which reduces to the conventional Zak phase expression in the linear limit $\frac{dh}{d\Sigma}\rightarrow 0$. \GJ{It is remarkable that the nonlinear Zak phase introduced here can be expressed as a single $k$-integral involving the kernal $\mathcal{K}(k)$.  That is, the conventional Zak phase and the nonlinear Zak phase can be respectively obtained by excluding or including the kernal $\mathcal{K}(k)$.}

\subsection{Iterative approach to real-space solutions under OBC}
\label{subsection:IterativeMethod}
 
The previous subsection on nonlinear Zak phase is one major tool we adopt to investigate momentum-space features. For real-space solutions, especially when the system is under OBC, we can only find the real-space solutions by brute-force computational tools. 
To complete our methodology description, we briefly describe here an iterative approach. For a nonlinear (state-dependent) Hamiltonian $H_{\rm {OBC}}$, the iteration process from state $\ket{\Psi_n}$ to state $\ket{\Psi_{n+1}}$ is as follows:

\begin{itemize}
    \item Compute $H_n = H_{OBC}(\ket{\Psi_n})$, the \CH{nonlinear state-dependent} Hamiltonian of the system under OBC, \CH{evaluated at }%associated with 
		the state $\ket{\Psi_n}$.
    \item Solve $H_n$ for its eigenstates $\ket{\Phi_i}$ with $i = 1,...,2N$. \GJ{Note that we have even number of lattice sites.}
    \item We then choose the new state $\ket{\Psi_{n+1}}$ as the \GJ{special eigenstate} $\ket{\Phi_i}$ closest in distance to the previous $\ket{\Psi_n}$, i.e. the state which minimizes $\left\| \ket{\Psi_n}-\ket{\Phi_i}\right\|$, where we have defined the norm $\left\|\ket{\psi}\right\|=\left|\langle \psi | \psi\rangle \right|$. In other words, $\ket{\Psi_{n+1}} = \ket{\Phi_{i_0}}$ where $\left\|\ket{\Psi_n}-\ket{\Phi_{i_0}}\right\| \leq  \left\|\ket{\Psi_n}-\ket{\Phi_i}\right\|$ for all $i$.
\end{itemize}

To execute the above-described iteration method, one also needs to choose the starting point of the iteration. In our studies, we choose them to be the initial states of the bulk eigenstates and the edge states of our model in the linear limit. We then iterate until the distance between old and new state is less than an arbitrary $\epsilon$, i.e. $\left\|\ket{\Psi_{n}}-\ket{\Psi_{n+1}}\right\| < \epsilon$. Throughout this work, we take $\epsilon = 10^{-10}$. \TT{Since the aforementioned iterative approach can only capture a limited number of stable stationary state solutions, and many choices of trial initial states may converge to the same state, we select only a subset of representative bulk eigenstates of the underlying linear model that converge to distinct solutions for numerical efficiency to obtain the energy spectra shown in Fig.~\ref{fig:EnergySpectrumG} and Fig.~\ref{fig:EnergySpectrumGTrivial}, and the inverse participation ratios shown in Fig.~\ref{fig:IPRofG}.}

\section{Nonlinear SSH model}
\label{section:NLSSHmodel}

%To explicitly demonstrate how nonlinearity suppresses the loss of topological protection due to chiral symmetry breaking perturbations, 
This work focuses on a nonlinear SSH chain of $N$ dimers as a case study. 
%most features observed below also exist for other more general models. 
Such a model is described by the following set of nonlinear Schr\"{o}dinger equations, %. We consider in addition a chiral-symmetry breaking term $v$ and an on-site nonlinearity of coefficient $g$ proportional to the occupation of the site. The coefficients $\Psi_{A,j}$ and $\Psi_{B,j}$ respectively denote the sites A and B of the $j$th cell. This system is described in real space by the nonlinear Schrödinger equations
\begin{equation}
\label{eqn:RealSpaceModel}
\begin{aligned}
i \frac{d \Psi_{A,j}}{dt} = J_1 \Psi_{B,j} + J_2 \Psi_{B,j-1} + v \Psi_{A,j} + g \left|\Psi_{A,j}\right|^2 \Psi_{A,j}\\
i \frac{d \Psi_{B,j}}{dt} = J_1 \Psi_{A,j} + J_2 \Psi_{A,j+1} - v \Psi_{B,j} + g \left|\Psi_{B,j}\right|^2 \Psi_{B,j}
\end{aligned}
\end{equation}
where $J_1$ and $J_2$ describe the intra- and inter-cell hopping amplitudes respectively, $v$ is a staggered onsite potential strength which breaks the system's chiral symmetry, $\Psi_{A,j}$ and $\Psi_{B,j}$ respectively denote the sites A and B of the $j$th cell, which satisfy $\Psi_{B,-1} = \Psi_{A,N+1} = 0$ under OBC or $\Psi_{A,N+1}=\Psi_{A,1}$ and $\Psi_{B,-1}=\Psi_{B,N}$ under PBC. 

In the linear limit, i.e., $g=0$, Eq.~(\ref{eqn:RealSpaceModel}) under PBC is governed by the momentum space Hamiltonian
\begin{equation}
H(k) = (J_1+J_2\cos(k))\sigma_x + J_2 \sin(k) \sigma_y + v\sigma_z \;. \label{linSSH}
\end{equation}
If $v=0$, which we will refer to as the unperturbed or chiral-symmetric SSH model in the rest of this paper, it satisfies $\Gamma H(k) \Gamma^\dagger = -H(k)$ with $\Gamma=\sigma_z$ being the chiral symmetry operator, which as discussed earlier leads to the quantized Zak phase $\gamma=\pi \frac{1+\mathrm{sgn}\left(J_2-J_1\right)}{2}\in\left\lbrace0,\pi \right\rbrace$. The case $\gamma=0$ ($\gamma=\pi$) corresponds to a topologically trivial (nontrivial) regime, where the system does not host (hosts) zero energy end states under OBC. That whether boundaries host edge states can be determined solely from the bulk properties represents an instance of the so-called bulk-boundary correspondence~\cite{BBC1,BBC2}. In this case, since $\gamma$ is only quantized to either $0$ or $\pi$ if the chiral symmetry is respected, the presence of a chiral-symmetry breaking term generally causes these edge states (if they exist) to lose their topological protection. In particular, taking $v\neq 0$ in Eq.~(\ref{linSSH}) in the regime \TT{$\gamma=\pi$} ($J_2>J_1$) leads to unequal shifts of the two end states to energy $\pm v$, so that one may then continuously tune $v$ to remove these edge states without closing the bulk energy gap. 

It should be highlighted that the nonlinear lattice model system depicted above is experimentally realizable in several existing experimental platforms. For example, within the framework of topological photonics~\cite{NonlinearDiracLikeSSHTopologialEdgeAndSoliton}, such a model can be realized by considering a one-dimensional (1D) array of waveguides, where each waveguide has unequal distances to its left and right adjacent waveguides so as to generate dimerized nearest-neighbor couplings $J_1$ and $J_2$ in the paraxial wave equation simulating Eq.~(\ref{eqn:RealSpaceModel}) above. A chiral-symmetry breaking term can be induced when waveguides with alternating refractive indices are arranged in the chain. Finally, on-site nonlinearity is naturally formed via the Kerr mechanism. Alternatively, the same model may also be qualitatively replicated with electrical circuit setups containing non-linear diodes~\cite{wang2019topologically,ExampleNonlinearProtectionPossiblePhaseTransition}.

In the following sections, we shall extensively study the role of nonlinearity in recovering some intriguing topological properties despite the chiral symmetry being broken. Representative results include a recovery of quantization regarding nonlinear Zak phases when PBC are applied and a recovery of degenerate edge states under OBC.

%In the standard, chiral symmetric SSH model, the bulk-boundary correspondence connects the quantized Zak phase under periodic boundary conditions (PBC) to the number of edge states of the system under open boundary conditions (OBC). As chiral symmetry is broken, the Zak phases associated to the energy bands are no longer quantized, and the degeneracy of the edge states is lifted. In this section, we will show that a good choice of nonlinear diagonal term makes possible the recovery of both properties. For the rest of this paper, we go back to considering the model of Eq.(\ref{eqn:RealSpaceModel}).

\section{Momentum-space Results}
\label{section:PBC}

In this section we investigate our nonlinear SSH model under PBC, where a nonlinear Hamiltonian of the form Eq.~(\ref{eqn:GPHamiltonian}) can be obtained from Eq.~(\ref{eqn:RealSpaceModel}) by further assuming Bloch state solutions 
\begin{equation}
\label{eqn:PhaseSpaceModel}
\begin{aligned}
\Psi_{A,j} = \Phi_{A} e^{ikj}\\
\Psi_{B,j} = \Phi_{B} e^{ikj},
\end{aligned}
\end{equation}
which gives us the nonlinear eigenvalue problem $H(\Sigma)\Phi = E \Phi$, with the pseudo-spinor $\Phi = [\Phi_A,\Phi_B]^{T}$ and a two-band Gross–Pitaevskii (GP) Hamiltonian
\begin{equation}
\label{eqn:BlochSpaceHamiltonian}
H(\Sigma) = (J_1 + J_2 \cos{k})\sigma_x + J_2 \sin{k} \sigma_y + h(\Sigma) \sigma_z + \frac{g}{2} I_2,
\end{equation}
where $h(\Sigma)=v + \frac{g}{2} \Sigma$, $\Sigma = \left|\Phi_{B}\right|^2 - \left|\Phi_{A}\right|^2$ is the population difference between the two pseudo-spinor components, \GJ{$I_2$ is a $2\times 2$ identity matrix}, and $\sigma_{x,y,z}$ are the Pauli matrices acting on the $\left[\Phi_A, \Phi_B\right]$ basis. 

\subsection{Nonlinear band structure}
\begin{figure}
  \includegraphics[width=\linewidth]{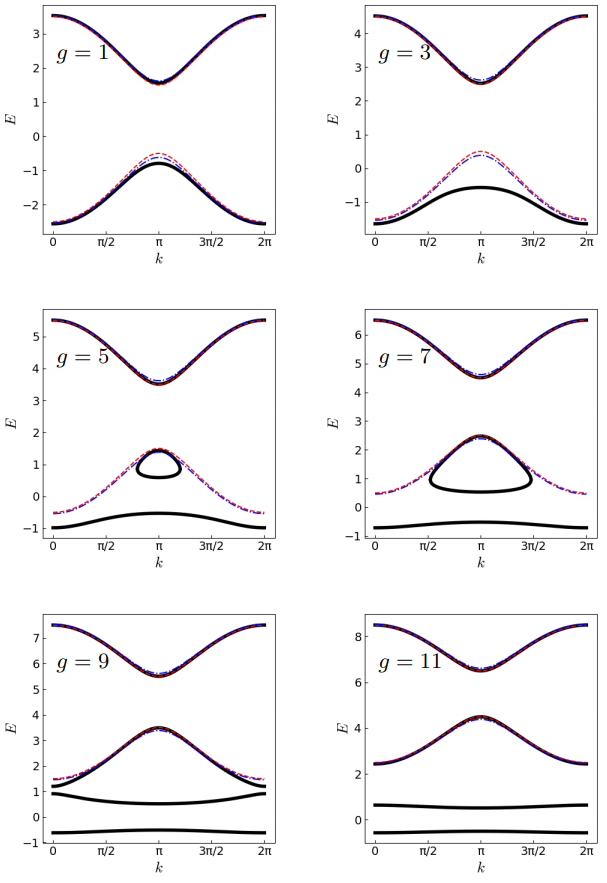}
  \caption{The energy bands of nonlinear chiral-symmetry broken SSH model vs nonlinearity strength $g$ (values of $g$ are indicated on figure sub-panels). The red dashed lines indicate the energy bands of the associated chiral-symmetric linear SSH model, described by the Hamiltonian of Eq.~(\ref{eqn:BlochSpaceHamiltonian}) with $h(\Sigma)=0$. The blue dash-dotted lines depict the energy bands of the associated linear SSH model with chiral symmetry broken, i.e., $h(\Sigma)=v$. The black continuous lines represent the energy bands of the nonlinear SSH model with both chiral symmetry breaking and Kerr-like nonlinearity, i.e., $h(\Sigma)=v+\frac{g}{2}\Sigma$. All quantities shown in the pictures are given in units of $J_1$, with parameter values $J_2=2 $, and $v=0.5 $.}
  \label{fig:EnergyBandsComparisonDifferentg}
\end{figure}

In Fig.~\ref{fig:EnergyBandsComparisonDifferentg}, we show the system's band structures vs  nonlinearity strength, and compare them with the energy bands of the associated SSH model with $g=0$, with and without chiral-symmetry breaking.
As the nonlinearity strength increases, a ``looped" band structure eventually emerges, which corresponds to additional energy bands that exist only within some region in the Brillouin zone, as depicted in Fig.~\ref{fig:EnergyBandsComparisonDifferentg} for $g=5$ and $g=7$. The region in the Brillouin zone for which these additional bands exist enlarges as $g$ increases, eventually spanning the entire Brillouin zone at large enough nonlinearity, as Fig.~\ref{fig:EnergyBandsComparisonDifferentg} shows for $g=9$ and $g=11$. That is, at very large values of $g$, four well defined energy bands exist in the system, two of which being really close to the bands of the chiral-symmetric linear SSH model. \GJ{To understand this, note that as $g\gg v$}, the Hamiltonian is approximately 
\begin{equation}
    \label{eqn:Large_g_Hamiltonian}
    H \approx  \begin{pmatrix} \frac{g}{2} + \frac{g}{2} \Sigma & J_1 + J_2 e^{-ik} \\ J_1 + J_2 e^{ik} & \frac{g}{2} - \frac{g}{2} \Sigma \end{pmatrix},
\end{equation}
\GJ{which allows for two eigenstates satisfying $\left|\Phi_2^{(0)}\right|^2 = \left|\Phi_1^{(0)}\right|^2$. These two eigenstates then cancel the nonlinear term and hence coincide precisely with that of the chiral-symmetric SSH model}. Thus, in spite of a chiral-symmetry breaking term, these two nonlinear energy bands are in fact very close to the bands of the unperturbed linear SSH model, suggesting the possibility of nonlinearity induced recovery of some topological features \CH{originally defined in the linear limit.} 
\begin{figure}
  \includegraphics[width=\linewidth]{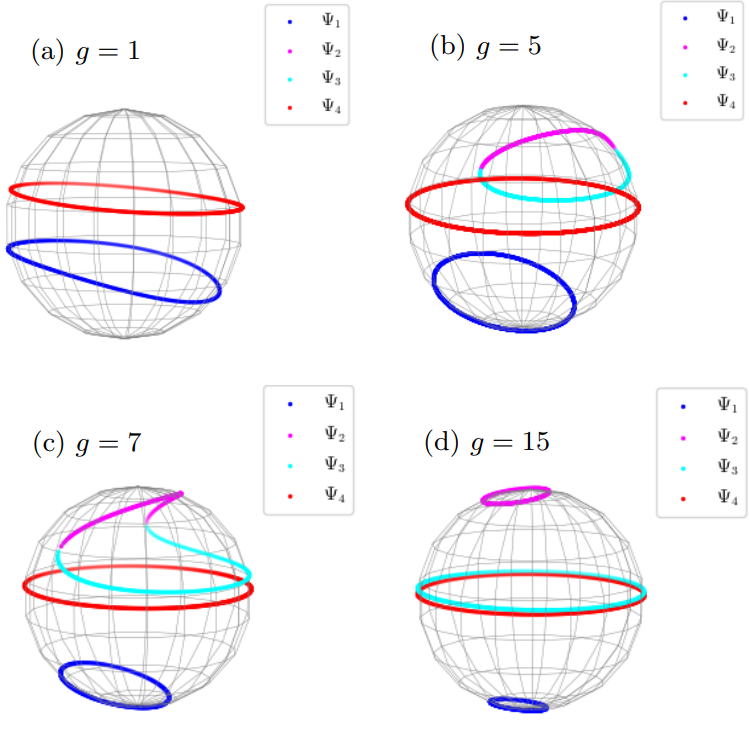}
  \caption{Bloch sphere representation of the adiabatic evolution of the system's stationary states $\Psi_i$ as $k$ is scanned over the Brillouin zone, with $E_1 < E_2 < E_3 < E_4$.  System parameters are $J_2=2 $, and $v=0.5 $ in units of $J_1$.}
  \label{fig:BlochSpheres}
\end{figure}

%we look at the solid angle drawn on the Bloch sphere by each stationary state as the quasimomentum $k$ runs over the Brillouin zone, as shown on Fig. \ref{fig:BlochSpheres}.

\subsection{Zak phase results}
\GJ{The conventional Zak phase reflects the geometrical path of adiabatic eigenstates. The geometric paths of the adiabatic eigenstates can be best shown in the Bloch sphere representation, for both linear and nonlinear SSH models}. To that end we first show in Fig.~\ref{fig:BlochSpheres} the Bloch sphere representation of the system's stationary states adiabatic evolution as the quasimomentum $k$ is scanned over the Brillouin zone. In the chiral-symmetric linear SSH model, the evolution of these states forms a closed loop along the equator of the Bloch sphere, which corresponds to a quantized $\pi$ Zak phase. In the presence of a chiral-symmetry breaking term, such a loop is deformed away from the equator, thus breaking the quantization of the Zak phase. This feature persists in the presence of weak nonlinearity, as depicted in Fig.~\ref{fig:BlochSpheres}(a). Remarkably, as the strength of nonlinearity continues to increase, one of these loops tends to move back towards the equator, while the other moves even farther away. As the looped band structure emerges and enlarges to become two additional energy bands, they individually trace out a closed loop on the Bloch sphere, which further approaches the equator as the nonlinearity strength further increases (see Figs.~\ref{fig:BlochSpheres}(b,c,d)). At very large nonlinearity strength, there are thus two nonlinear bands with almost $\pi$ quantized Zak phase. These two bands are precisely those observed in Fig~\ref{fig:EnergyBandsComparisonDifferentg} at $g=9$ and $g=11$ that closely resemble the two bands of the unperturbed linear SSH model. %This, combined with the similarity of the energy bands to the ones of a chiral symmetric SSH model hints for a possible quantization of the nonlinear Zak phase of the nonlinear energy bands. Using a numerical scheme for the integration of the Zak phase \cite{NumericalSchemeForNLBP} we adapt it to compute the nonlinear Zak phase as the nonlinearity varies.

\begin{figure}
  \includegraphics[width=\linewidth]{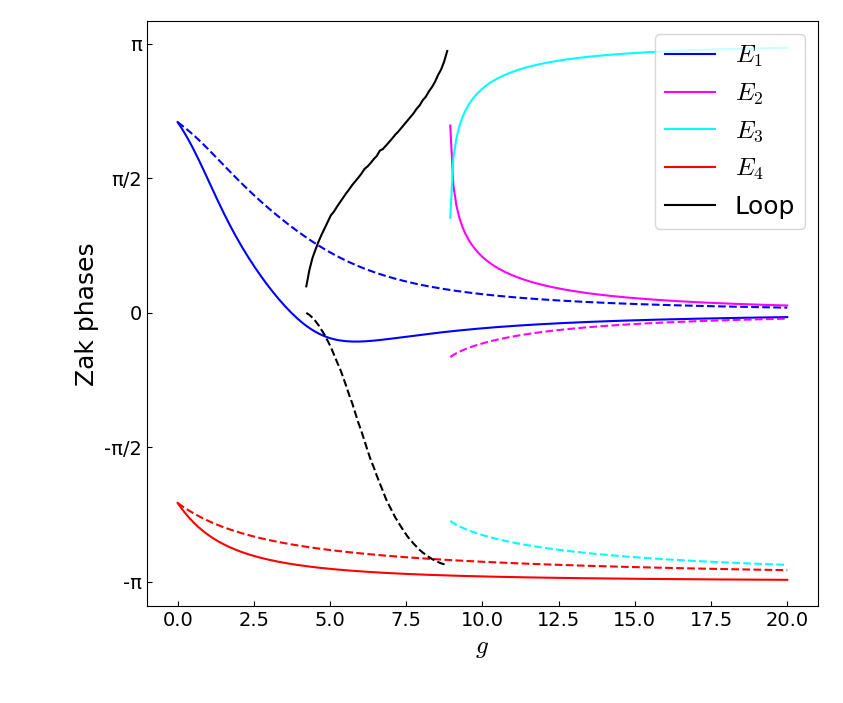}
  \caption{Conventional Zak phases and nonlinear Zak phases of each band vs nonlinear strength. The continuous (dashed) lines represent nonlinear (conventional) Zak phases computed by including (excluding) the deforming kernel $\mathcal{K}$ in Eq.~(\ref{eqn:CorrectionNLBP}). Note that each Zak phase converges to a quantized value of either $0$ or $\pi$ in the large $g$ limit. The Berry phase of one cycle around the looped structure is also included.  System parameter values are $J_2=2$, and $v=0.5$, in units of $J_1$.}
  \label{fig:IndividualZakPhases}
\end{figure}

\GJ{One may wonder how the concept of nonlinear Zak phase introduced in Sec.~\ref{section:Computational tools} helps us to appreciate the physics here further.}  Let us now quantitatively examine  the nonlinear or conventional Zak phases associated with all the system's available energy bands, accomplished by adapting the scheme presented in Ref.~\cite{NumericalSchemeForNLBP}. The results are presented in Fig.~\ref{fig:IndividualZakPhases}, where Zak phases associated with the energy bands $E_1<E_2<E_3<E_4$, with and without the kernel $\mathcal{K}$ derived in Sec.~\ref{section:Nonlinear Zak phase}, are plotted vs nonlinearity strength $g$.  Regarding the looped band structures that represent two incomplete energy bands, they together form a closed loop on the the Bloch sphere representation. Hence it is also of some interest to evaluate the Berry phase associated with this peculiar looped structure when it exists.  This is done by scanning the system from the smallest quasimomentum for which the incomplete band exists, going all the way to the other extremity of the incomplete band, before coming back to the starting point by scanning the other incomplete energy band, thus performing a closed path.

\begin{figure}
    \centering
    \includegraphics[width=\linewidth]{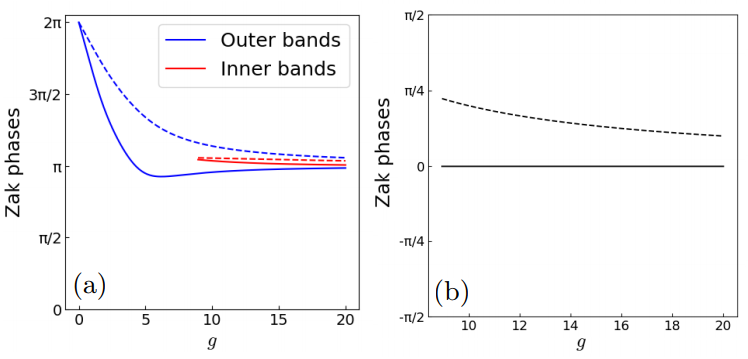}
    \caption{Panel (a): Sum of the Zak phases of the two outermost energy bands $E_1$ and $E_4$ and of the two innermost energy bands $E_2$ and $E_3$. Panel (b): Sum of the Zak phases of all four energy bands. For both panels, the continuous (dashed) lines represent the nonlinear (conventional) Zak phases computed with (without) the kernel $\mathcal{K}$ in Eq.~(\ref{eqn:CorrectionNLBP}). System parameters are $J_2=2 J_1$, and $v=0.5 J_1$.}
    \label{fig:SumZakPhases}
\end{figure}

\GJ{Our main findings from Fig.~\ref{fig:IndividualZakPhases} are as follows. As nonlinearity strength increases, the nonlinear or the conventional Zak phases of energy bands $E_1$ and $E_4$, which resemble those of the chiral-symmetric linear SSH model, become closer to $\pi$, though they are never exactly quantized. It is further observed that the nonlinear Zak phases can be significantly different from the conventional Zak phases. In particular, the nonlinear Zak phases for bands $E_1$ and $E_4$ at large nonlinearity strength are closer to a quantized $\pi$ value than the conventional Zak phases.  More importantly, an exact quantization of the summation over the four nonlinear Zak phases at $0$ modulo $2\pi$ is recovered in the regime where four well-defined energy bands exist, as shown in Fig.~\ref{fig:SumZakPhases}(b). This quantization is broken at low nonlinearity strength, due to the presence of incomplete energy bands. It is also interesting to notice that the Berry phase associated with the looped band structure gradually changes from $0$ to $\pi$, as the peculiar loop band structure first emerges and disappears at large nonlinear strength.}

\GJ{To better understand the recovery of quantization of the summation of all nonlinear Zak phases at $0$ modulo $2\pi$}, we have also applied a perturbation theory to obtain approximate expressions for the four nonlinear or conventional Zak phases for large nonlinearity strength. As further detailed in Appendix~\ref{app2}, by treating $\frac{1}{g}$ as a perturbative parameter and making simplifying assumptions that $J_1=0$, the nonlinear Zak phases $\gamma_1$, $\gamma_2$, $\gamma_3$ and $\gamma_4$ associated with energy bands $E_1<E_2<E_3<E_4$ are found to be %This quantized quantity is inherent to the nonlinear adiabatic perturbation theory used to determine the deforming kernel to apply to the Berry curvature, as the same quantity is non longer quantized if computed with only the standard Berry curvature. Finally, we confirm this result using perturbation theory. We first considered the chiral symmetry breaking term $v \sigma_z$ in \ref{eqn:BlochSpaceHamiltonian} as a small perturbation, to obtain the first order corrections to the eigenenergies and eigenstates of the four bands. We then made the additional assumptions that $J_1=0$ and $J_2 \ll g$ in order to be able to analytically compute the correction to the Zak phases. Writing $\gamma_1$, $\gamma_2$, $\gamma_3$ and $\gamma_4$ the Zak phase (with nonlinear correction) associated to the energy bands $E_1<E_2<E_3<E_4$, we get 
\begin{equation}
	\label{eqn:PerturbativeZakPhase}
    \begin{aligned}
        \gamma_1 &= - 2\pi \left(\frac{J_2}{g}\right)^2 + \mathcal{O}\left(\frac{v J_2^2}{g^3}\right)\\
        \gamma_2 &= 2 \pi \left(\frac{J_2}{g}\right)^2 + \mathcal{O}\left(\frac{v J_2^2}{g^3}\right)\\
        \gamma_3 &= -\pi \left( 1 +  \frac{4 v J_2}{g^2}\right) + \mathcal{O}\left(\frac{v J_2^2}{g^3}\right)\\
        \gamma_4 &= -\pi \left( 1 -  \frac{4 v J_2}{g^2}\right) + \mathcal{O}\left(\frac{v J_2^2}{g^3}\right).
    \end{aligned}
    \end{equation}
\GJ{Clearly, the sum of these four nonlinear Zak phases is quantized.  This further confirms our computational findings, though our computational findings are valid to higher orders of $1/g$.}   
\GJ{By contrast,, the four conventional Zak phases $\gamma_1'$, $\gamma_2'$, $\gamma_3'$ and $\gamma_4'$ associated with the same energy bands (that is, excluding geometric phase contributions from the Kernel $\mathcal{K}(k)$) are obtained as}
\begin{equation}
    \label{eqn:PerturbativeLinearZakPhase}
    \begin{aligned}
        \gamma_1' &= 2\pi \left( \frac{J_2}{g} \right)^2 + \mathcal{O}\left(\frac{v J_2^2}{g^3} \right), \\
        \gamma_2' &= - 2 \pi \left( \frac{J_2}{g} \right)^2 + \mathcal{O}\left(\frac{v J_2^2}{g^3} \right), \\
        \gamma_3' &= -\pi (1 - 2 \frac{v}{g} - 4 \frac{v J_2}{g^2} ) + \mathcal{O}\left(\frac{v J_2^2}{g^3}, \right) \\
        \gamma_4' &= - \pi (1 - 2 \frac{v}{g} + 4 \frac{v J_2}{g^2} ) + \mathcal{O}\left(\frac{v J_2^2}{g^3}.\right)
    \end{aligned}
 \end{equation} %\blue{RWB: Hopefully this additional calculation can also be done, which I think is important to highlight how the nonlinear correction factor plays a role in making the sum of four Zak phases quantized.} %The first correction to the Zak phases of the two near-SSH energy bands is of order $\frac{v J_2}{g^2}$ and the sum of all four Zak phases is quantized up to $\mathcal{O}\left(\frac{v J_2^2}{g^3}\right)$. The details  can be found in Appendix~\ref{app2}. 
\GJ{The sum of these conventional Zak phases is clearly not quantized. That only nonlinear Zak phases may recover quantization is a remarkable observation. This finding also echoes with an early study by two of the present authors and others~\cite{NonlinearDiracCones}, where it was found that only a nonlinearity corrected Aharonov-Bohm phase is quantized around nonlinear Dirac cones. Topological characterization of nonlinear lattice systems hence has unique features absent in linear systems.} 

\subsection{Dynamical stability of solutions}
We will now investigate the dynamical stability of the obtained nonlinear energy bands above. To this end, we evaluate the time evolution of a state initially prepared slightly away from a stationary state, assuming for simplicity that such a state also respects the translational symmetry of the system, which is obtained by solving the time dependent GP equation in the momentum space
\begin{equation}
    \label{eqn:DynamicalPerturbation}
    i \frac{\partial}{\partial t} \ket{\Psi(k,t)} = H(k,\Psi(k,t)) \ket{\Psi(k,t)},
\end{equation}
where $H$ is given by Eq.(\ref{eqn:BlochSpaceHamiltonian}). Such a state $\ket{\Psi(k,t)}$ can be written as a sum of a stationary state $\ket{\psi(k,t)}$ with energy $E(k)$ and a small perturbation of the form 
\begin{equation}
\label{eqn:ComponentsPerturbation}
    \ket{\delta\psi(k,t)}=\begin{pmatrix} \delta\psi_1(k,t) \\ \delta\psi_2(k,t) \end{pmatrix}.
\end{equation}
We then define the stationary solution $\ket{\psi(k,t)}$ to be dynamically stable if \CH{the norm of} $\ket{\Psi(k,t)}= \ket{\psi(k,t)}+ \ket{\delta\psi(k,t)}$ does not go to $\infty$ as $t \rightarrow \infty$ for sufficiently small $\ket{\delta\psi(k,0)}$. For clearer calculations, we separate the "dynamical phase" from $\ket{\Psi(k,t)}$ as
\begin{equation}
\label{eqn:SeparateDynamicalPhase}
    \begin{aligned}
    \ket{\Psi(k,t)} &= e^{-iEt} \ket{\Phi(k,t)}, \\
    \ket{\psi(k,t)} &= e^{-iEt} \ket{\psi(k,0)}, \\
    \ket{\delta \psi(k,t)} &= e^{-iEt} \ket{\delta \phi(k,t)},
    \end{aligned}
\end{equation}
\CH{so as to form a resultant state $\ket{\Phi(k,t)}=\ket{\psi(k,0)}+\ket{\delta \phi(k,t)}$ that separates} into a time independent part $\ket{\psi(k,0)}$ plus a time dependent part $\ket{\delta \phi(k,t)}$. With some algebra, Eq.(\ref{eqn:DynamicalPerturbation}) can be written in the following form, 
\begin{equation}
    \label{eqn:RecastDynamicalPerturbation}
    i \frac{\partial}{\partial t} \begin{pmatrix} \delta\phi_1 \\ \delta\phi_2 \\ \delta\phi_1^{*} \\ \delta\phi_2^{*} \end{pmatrix} = \mathcal{L} \begin{pmatrix} \delta\phi_1 \\ \delta\phi_2 \\ \delta\phi_1^{*} \\ \delta\phi_2^{*} \end{pmatrix},
\end{equation}
where
\begin{equation}
    \label{eqn:MatrixDynamicalEvolution}
    \begin{aligned}
    \mathcal{L} &=  \begin{bmatrix} H_{gp} + A & B  \\\ -B^{\dagger} & -H_{gp} - A^{*} \end{bmatrix}, \\
    H_{gp} &= H(k,\psi(k,0)) - E I_2, \\
    A &= \frac{g}{2} \begin{pmatrix} -|\psi_1(k,0)|^2 & \psi_1(k,0) \psi_2(k,0)^{*} \\\ \psi_1(k,0)^{*} \psi_2(k,0) & -|\psi_2(k,0)|^2 \end{pmatrix}, \\
    B &= \frac{g}{2} \begin{pmatrix} -\psi_1(k,0)^{2} & \psi_1(k,0) \psi_2(k,0) \\\ \psi_1(k,0) \psi_2(k,0) & -\psi_2(k,0)^{2} \end{pmatrix}.
    \end{aligned}
\end{equation}
As Eq.~(\ref{eqn:RecastDynamicalPerturbation}) resembles the time dependent Schrödinger equation in linear quantum mechanics, its time evolution is governed by the operator $e^{-i \mathcal{L} t}$. However, since $\mathcal{L}$ is not a Hermitian operator, eigenvalues of $\mathcal{L}$ can in general be complex. It follows that in order for $\ket{\phi(k,t)}$ to be dynamically stable, all eigenvalues $\lambda_n$ of $\mathcal{L}$ must satisfy \cite{DynamicalStabilityCondition},
\begin{equation}
    \label{eqn:DynamicalStabilityCondition}
    \operatorname{Im}(\lambda_n) = 0 , \forall n .
\end{equation}
\begin{figure}
  \centering
  \includegraphics[width=\linewidth]{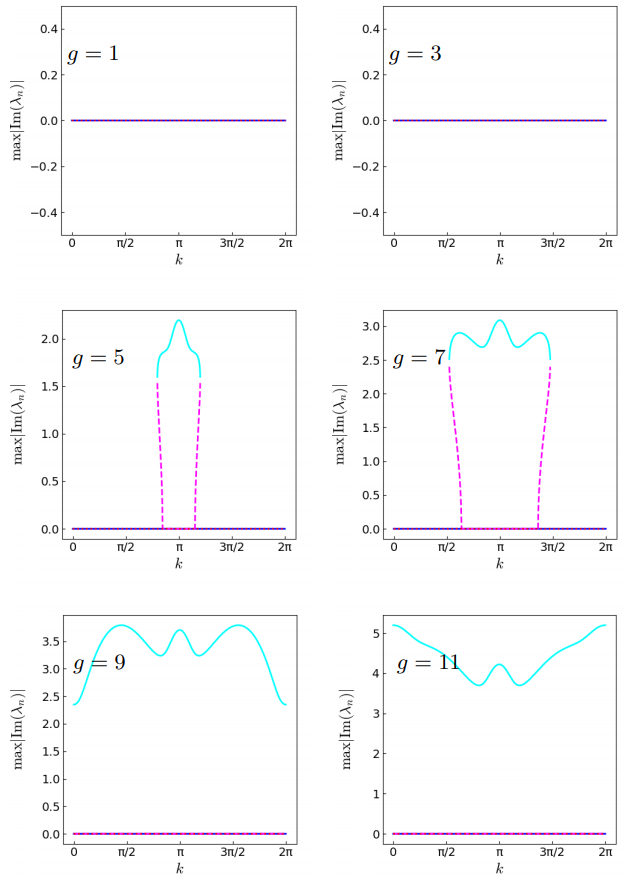}
  \caption{$\max |\operatorname{Im}(\lambda_n)|$ for the nonlinear energy bands $E_1 < E_2 < E_3 <E_4$. The blue, magenta, cyan, and red lines correspond to $E_1$, $E_2$, $E_3$ and $E_4$ respectively. System parameters are $J_2=2 $, and $v=0.5 $ in units of $J_1$.}
  \label{fig:DynamicalStabilityDifferentg}
\end{figure}

Figure~\ref{fig:DynamicalStabilityDifferentg} shows the maximum imaginary component of all the eigenvalues of $\mathcal{L}$ for all the nonlinear bands, from which it follows that the highest and lowest energy bands, the former being the band that closely resembles chiral-symmetric linear SSH model and possesses an almost $\pi$ quantized Zak phase at large nonlinearity strengths, are dynamically stable throughout the Brillouin zone. On the other hand, the second largest band $E_3$ is dynamically unstable whenever it exists, whereas the second lowest band $E_2$ shows \CH{in}stability for some values of $k$ when the looped structure exists, which becomes fully stable once the nonlinearity strength is large enough for four complete bands to exist. 

%In this model, the operator $\mathcal{L}$

%\textcolor{red}{[CH: Need to mention the significance of these results. Does complex eigenvalues lead to instabilities that break the bulk boundary correspondence? If yes, only for certain $k$ values? What about the k-points where even one of the inner bands have real spectrum?\\In my opinion, this stability analysis with $\mathcal{L}$ is one of the most important keys of understanding what happens when non-linearity interplays with the change in boundary conditions. In particular, can the lack of the bulk boundary correspondence be understood via the non-hermitian skin effect? This seems very interesting.... ]}

\section{Real-space results}
\label{section:OBC}

\subsection{Spectrum and eigenstates under OBC}
\begin{figure}
  \centering
  \includegraphics[width=\linewidth]{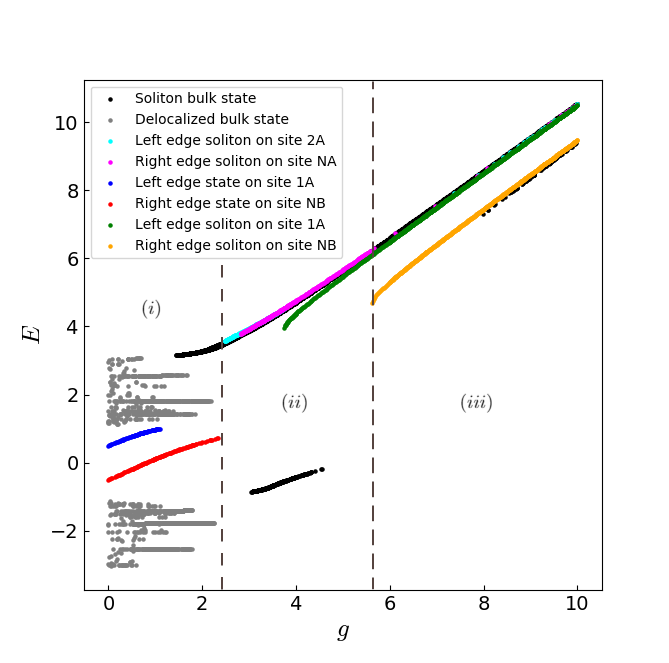}
  \caption{Energy spectrum solved from Eq.~(\ref{eqn:RealSpaceModel}) under OBC, showing three different regimes $(i)$, $(ii)$ and $(iii)$. All quantities shown are given in units of $J_1$, with parameter values $J_2=2 $, $v=0.5 $, and $N=100$ unit cells. Under these parameters, the associated chiral-symmetric SSH model with $g=0$ would be in the topological nontrivial regime. }
  \label{fig:EnergySpectrumG}
\end{figure}

We now shift our focus to the real-space behavior of nonlinear lattice systems under OBC, using again the model described by Eq~(\ref{eqn:RealSpaceModel}).  Computationally we use the iterative method already introduced in Sec.~\ref{section:Computational tools},  taking both the bulk eigenstates and edge states of the chiral-symmetry broken SSH model as initial trial states.  We then numerically obtain the energy spectrum under OBC, for different values of $g$, as shown in Fig.~\ref{fig:EnergySpectrumG}.

\begin{figure}
    \centering
    \includegraphics[width=\linewidth]{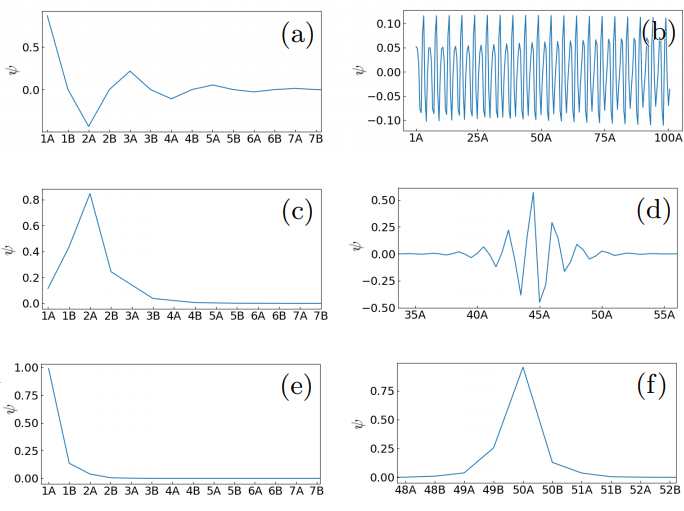}
    \caption{Wave function profiles of different types of states existing in each regime, with system parameters given by $J_2=2$, and $v=0.5$ in units of $J_1.$
Panels (a) illustrates one edge state on site $1A$ for $g=1$, 
(b) one delocalized bulk state for $g=1$,
(c) one edge soliton localized at site $2A$ for $g=3.5$,
 (d) one in-gap soliton solution for $g=3.5$,
 (e) one edge soliton localized at site $1A$ for $g=7$,
and (f) one bulk soliton state for $g=7$. \TT{A value of $g=1$ places the system in regime $(i)$, $g=3.5$ in regime $(ii)$ and $g=7$ in regime $(iii)$.}\\}
    \label{fig:TypesOfStates}
\end{figure}

\begin{figure}
    \centering
    \includegraphics[width=\linewidth]{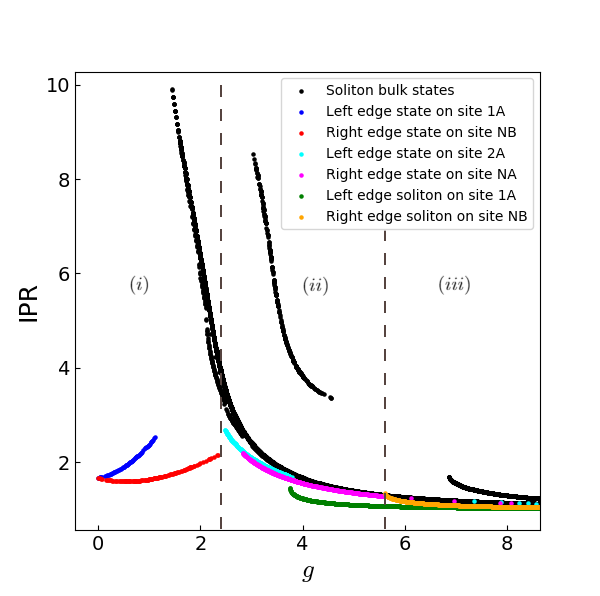}
    \caption{Inverse participation ratios of different types of localized states in the three nonlinearity regimes for system parameters in the topological nontrivial regime of the associated chiral-symmetric linear SSH model. The delocalized bulk states are not shown, as their IPR is greater than 100. System parameter values are $J_2=2 J_1$, and $v=0.5 J_1$.} 
    \label{fig:IPRofG}
\end{figure}

We may separate the typical energy spectrum as depicted in Fig.~\ref{fig:EnergySpectrumG} into there different regimes, depending on the strength of nonlinearity. Each regime accommodates different types of states, which are presented in Fig.~\ref{fig:TypesOfStates}.
As the nonlinearity strength increases, we observe a progressive break down of the energy bands obtained under PBC, as delocalized states disappear and are replaced by soliton states. \TT{Here, we use the term ``soliton" loosely, to refer to any localized state that is not directly related to an edge state of the model in the linear limit. This applies to any state existing in regime $(ii)$ and $(iii)$ in Fig.~\ref{fig:EnergySpectrumG}, where nonlinearity plays a substantial role.} In the low nonlinearity regime $(i)$, the original two bands of the linear SSH model remain occupied by delocalized bulk states such as the ones shown in Fig.~\ref{fig:TypesOfStates}(b).  We also observe two edge states localized at sites $1A$ and $NB$ (cf Fig.~\ref{fig:TypesOfStates}(a)), and these two edge states are nondegenerate due to the chiral symmetry breaking term we introduced to the system. However, these edge states are still very much akin to the ones of the linear case, as the nonlinearity is still too weak to destroy them. \TT{The disappearance of the last edge state marks the end of the low nonlinearity regime $(i)$}. On the other hand, if we consider the strong nonlinearity regime $(iii)$, where the nonlinearity is dominant over other energy scales, the only type of states that can be observed are two highly degenerate, large-energy solitons, located at any single site in the bulk (e.g. 
Fig.~\ref{fig:TypesOfStates}(f)) or at an edge (e.g. Fig.~\ref{fig:TypesOfStates}(c)). These solitons are non-topological, as they are simply the consequence of nonlinearity strength $g$ being much larger than all other energy scales of the system, They are related to the trivial single-site solutions in the limit $g \rightarrow \infty$, where all nonlinear eigenstates are exactly supported by only a single site (whose energy depends on whether sublattice A or B is occupied). This understanding is further supported by studying the inverse participation ratio (IPR) of the states, as show in Fig.~\ref{fig:IPRofG}. 
The IPR of a state $\ket{\Psi}$ is defined by 
\begin{equation}
\label{IPR_Def}
    \operatorname{IPR}(\ket{\Psi}) = \frac{1}{\sum\limits_{n=1}^{N} |\Psi_n|^4},
\end{equation}
and is small for localized states, but large ($\sim N$) for bulk delocalized states. It is seen that as the nonlinearity strength becomes large, all the nonlinear eigenstates become more and more localized, going towards an IPR of 1 (supported by a single site) as $g \rightarrow +\infty$. \TT{The existence of non-topological solitons on both edges of the system delimits the boundary of the strong nonlinearity regime $(iii)$}

\subsection{Soliton solutions with topological origin}
There is, however, an intermediate regime $(ii)$ of nonlinearity strength, where both nonlinear effects and topological properties of the linear model become important. This unique interplay between nonlinearity and topology can be understood by studying a special kind of bulk solitons whose energy is in the gap between the original linear energy bands (the energy of the edge states in regime $(i)$ is also in the gap). The profile of one such gap soliton is shown in Fig.~\ref{fig:TypesOfStates}(d). This profile indicates that on two respective sides of the soliton peak, there are two edge states emerging due to this effective nonlinearity induced ``edge" inside the bulk. This insight of an ``effective edge" in the bulk can be one main feature through which nonlinearity and topology can work conjointly in the system. That is, because the Hamiltonian here depends on the state, a wave function strongly localized at one site increases the potential energy there, effectively creating a potential barrier, which can be a large on-site potential for strong nonlinearity strength, thus effectively behaving like a physical edge. In turn, as fingerprints of the underlying topological phase of the associated chiral-symmetric SSH model, such an effective edge admits a strongly localized wave function, whose probability density exponentially decays with the distance from this effective edge.  These two feedback mechanisms thus allow such solitons of a topological origin to exist self-consistently. This understanding makes it clear that the existence of these gap solitons relies heavily on some recovered  topological features and represents a new, fascinating example of interplay between nonlinearity and topology. 

\begin{figure}
    \centering
    \includegraphics[width=\linewidth]{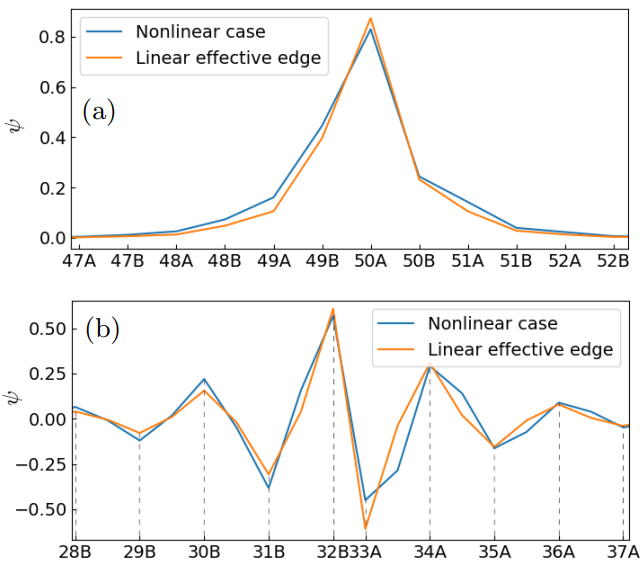}
    \caption{Comparison of various soliton solutions obtained from our nonlinear models with the corresponding localized eigenstates in the linear chiral-symmetric SSH model with added impurity potential as the edge in the bulk In the linear model, all solutions are either the two edge states due to the impurity potential or delocalized bulk states. System parameters are $J_2=2 $, $v=0.5 $, and $g=3.5$ in units of $J_1$.
Panel (a) illustrates one soliton solution of the nonlinear model vs a localized eigenstate in the linear unperturbed model with a single-site impurity-potential on site 50A. 
%Panel (b) shows the same results, but plotted in the log scale. 
Panel (b) compares another type of (in-gap) soliton solutions obtained from our nonlinear model with one localized eigenstate in the linear chiral-symmetric model with two impurity potential of  height $\frac{g}{2}$ introduced on sites $32B$ and $33A$.}
    \label{fig:CompareLinear}
\end{figure}

To further confirm that the peculiar soliton soution profiles can be understood as a combination of an effective edge and topological edge states, we now compare the in-gap soliton solutions with states in a linear chiral-symmetric SSH model plus an impurity in the bulk. Specifically, we consider then a chiral-symmetric linear SSH model with an additional impurity potential barrier of intensity $g$, placed only on one site in the system first (hence also playing the role of an effective edge inside the bulk). The model in real space can be described by 
\begin{equation}
\label{eqn:D_OBCEquations}
\begin{aligned}
i \frac{d \Psi_{A,j}}{dt} = J_1 \Psi_{B,j} + J_2 \Psi_{B,j-1} + v \Psi_{A,j} \textnormal{ if } j\neq j_0 \\
i \frac{d \Psi_{B,j}}{dt} = J_1 \Psi_{A,j} + J_2 \Psi_{A,j+1} - v \Psi_{B,j} \textnormal{ if } j\neq j_0
\end{aligned} \;,
\end{equation}
as well as 
\begin{equation}
\label{eqn:D_OneSiteEquation}
\begin{aligned}
i \frac{d \Psi_{A,j_0}}{dt} = J_1 \Psi_{B,j_0} + J_2 \Psi_{B,j_0-1} + (v+g) \Psi_{A,j_0} \\
i \frac{d \Psi_{B,j_0}}{dt} = J_1 \Psi_{A,j_0} + J_2 \Psi_{A,j_0 +1} + (-v+g) \Psi_{B,j_0}  
\end{aligned} \;,
\end{equation}
with $\Psi_{B,-1} = \Psi_{A,N+1} = 0$ under OBC. \GJ{We further set $v=0$ above for a linear chiral-symmetric SSH  model.} 
Remarkably, by setting the impurity potential at site $50A$ (we consider an example with $N=100$ unit cells) and solving for the eigenstates, we find one eigenstate highly resembling to one type of soliton solutions observed in our nonlinear model (see Fig.~\ref{fig:TypesOfStates}(f)).  This comparison is presented in Fig.~\ref{fig:CompareLinear}(a).
%In terms of the logarithm of each site's occupation probability for the high energy soliton, panel (b) of Fig~\ref{fig:D_LogBulkSolitonCompare} captures oscillations in the exponential decay behavior, favoring occupation probability on sites A on the right of the effective edge (impurity potential) and sites B on the left. 
%This strongly suggests that the above-observed soliton solutions have fingerprints of  topological properties %recovered by nonlinearity.  
%Other calculations also show that these oscillations become weaker as the nonlinearity strength increases and the %state then becomes trivially localized on one site in the limit $g \longrightarrow \infty$.

For another type of in-gap soliton solutions as illustrated in Fig.~\ref{fig:TypesOfStates}(d), the peaks of such soliton solutions are only localized on sublattice B on their left and sublattice A on their right  (assuming that our system are in the topologically non-trivial regime of the associated 
chiral-symmetric linear SSH model). This hence effectively creates two edges in the bulk with a new profile affecting the whereabouts of edge states. To confirm this understanding,  we accordingly introduce  two impurity potentials to the linear chiral-symmetric model, in the same manner as described above. The impurity potentials are of strength $\frac{g}{2}$ and are next to each other, the left one being on sublattice B and the right one on sublattice A. As shown Fig.~\ref{fig:CompareLinear}(b), we again obtain spatial profiles of localized states very close to the soliton solutions we found from the nonlinear model.

\begin{figure}
    \centering
    \includegraphics[width=\linewidth]{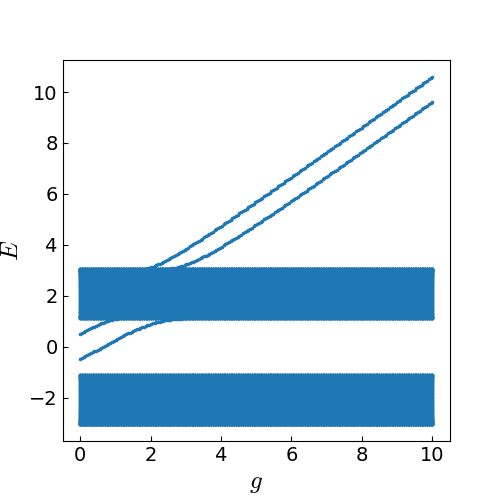}
    \caption{Energy spectrum of the linear SSH model with two potential barrier on sites $1A$ and $NB$, as a function of the strength $g$ of impurity potential introduced. All quantities shown are given in units of $J_1$, with parameter values $J_2=2 $ and $v=0.5$.}
    \label{fig:D_SprectrumDestroyEdges}
\end{figure}

Our impurity model can be also used to confirm that an effective edge on sites $1A$ or $NB$ indeed respectively destroys the existence of physical edge states on sites $1A$ or $NB$, with a topological explanation. We consider then two impurity potentials of strength $g$ added to sites $1A$ and $NB$, \TT{set the chiral symmetry breaking term $v$ to a nonzero value,} and then look into the energy spectrum of the linear system. The results in 
Fig.~\ref{fig:D_SprectrumDestroyEdges}
show that, as the strength of impurity increases, the two edge states are pushed away from zero energy until they merge with the bulk, after which they then disappear. Then, as the impurity strength further increases, two eigenstates with highest energy values are seen to emerge out of the bulk, a behavior akin to the edge solitons encountered in the high nonlinearity regime $(iii)$ seen above. \GJ{This clearly explains why there is an intermediate range of nonlinear strength $g$ for which edge solitons do not exist.} 

%It is further confirmed in Appendix~\ref{app4}, where the study of a linear SSH model, for the same given range $of parameter values, with an added potential barrier at one site of the bulk leads to the existence of very similar %localized eigenstates.

With the physical insights developed above, we are now ready to digest the recovery of the degeneracy of two edge solitons, in spite of the chiral-symmetry breaking term.  This important observation is presented in Fig.~\ref{fig:EnergySpectrumG} in the intermediate regime $(ii)$. There exists a range of $g$ values for which there are no edge solitons localized on the outermost sites $1A$ and $NB$. As sites $A$ and $B$ respectively bear the potential $+v$ and $-v$, a $2v$ energy difference exists between the states localized at these different sites, and this is the very reason why there is a splitting in the energy values of edge states localized at the very left or the very right. However, in this particular intermediate nonlinearity regime, the leftmost and rightmost localized states that do exist are respectively localized on sites $2A$ and $NA$ (e.g. 
Fig.~\ref{fig:TypesOfStates}(c)), effectively bypassing the energy splitting due to the broken chiral-symmetry (since they are both localized on sublattice $A$). \GJ{The absence of states localized on sites $1A$ and $NB$ can be traced back to a topological phase transition}. That is, the existence of a peak at site $1A$ leads to an additional nonlinearity induced edge potential at site 1A. The first site following such an edge is now a sublattice B, so the system now have alternating hopping amplitudes, acting again like an SSH model, but in the trivial regime because the roles of $J_1$ and $J_2$ have been exchanged, and hence cannot accommodate edge states. This hence indicates that such a soliton peaked at site 1A does not form a self-consistent solution to our nonlinear problem. A similar reasoning follows to arrive at the conclusion that a soliton localized on site $NB$ cannot self-consistently exist, either.  

\begin{figure}
  \centering
  \includegraphics[width=\linewidth]{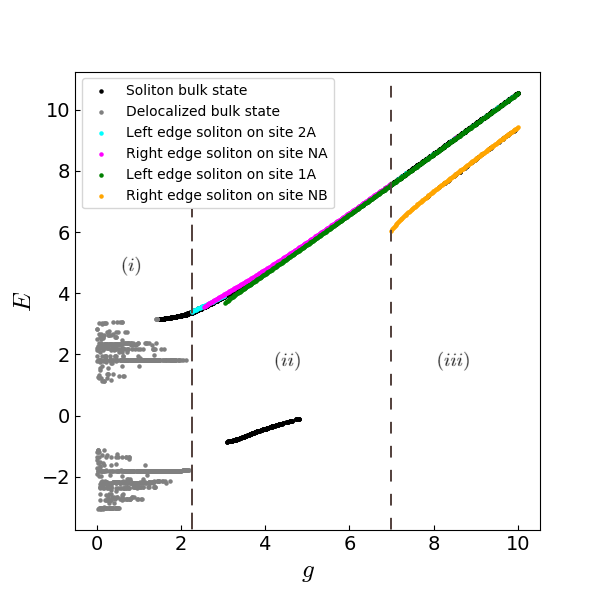}
  \caption{Energy spectrum of the originally topologically trivial ($J_1 > J_2$) model under OBC, which can be also divided into three regimes of nonlinearity strength. \TT{All quantities shown are in units of $J_2$, with parameter values $J_1=2 $,  $v=0.5 $}, and $N=100$ unit cells.}
  \label{fig:EnergySpectrumGTrivial}
\end{figure}

\begin{figure}
    \centering
    \includegraphics[width=\linewidth]{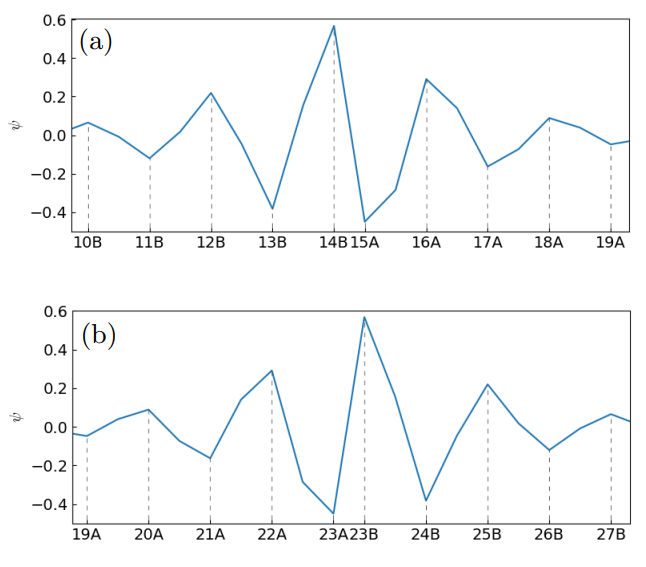}
    \caption{Wave function profiles of in-gap solitons when the associated chiral-symmetric linear system is in the topologically non-trivial and trivial regimes.  System parameters in panel (a) are $J_2=2 $, $v=0.5 $, and $g=3.5$ in units of $J_1$.  In panel (b),  the values of $J_1$ and $J_2$ are exchanged but leaving all other parameters unchanged, in order to connect with the topologically trivial case in the linear chiral-symmetric SSH model.}
    \label{fig:MirrorGapSolitons}
\end{figure}

As another remarkable consequence of the above intriguing mechanism effectively causing the exchange between the roles of $J_1$ and $J_2$, analogous soliton solutions can be expected at intermediate nonlinearity strength even when the corresponding linear system is in the topologically trivial regime. This is clearly evidenced in Fig.~\ref{fig:EnergySpectrumGTrivial}.
In particular, it is seen that even if the edge states originating from the linear model do not exist, in-gap solitons can be found. To confirm that this is indeed a consequence of the above-mentioned exchange between the roles of $J_1$ and $J_2$, we compare the in-gap soliton profiles between originally topologically trivial and non-trivial cases in  Fig~\ref{fig:MirrorGapSolitons}. Remarkably, %as clearly shown in  Fig~\ref{fig:MirrorGapSolitons}
 the respective typical soliton profiles from each case are mirror reflections of each other.  \GJ{Nonlinearity can thus not only recover topological properties destroyed by a chiral-symmetry breaking term, but also effectively induce topological features absent in the non-interacting limit.}

\section{Concluding Remarks}
\label{section:Conclusion}

In this work, we have carefully investigated the interplay between topology and nonlinearity in a simple SSH model with on-site nonlinearity and chiral symmetry breaking, with both momentum-space and real-space studies. The focus is on how nonlinearity may recover topological features analogous to a linear chiral-symmetric SSH model. We demonstrate that in the regime of strong nonlinearity, the nonlinear Zak phases (not the conventional Zak phases) of the nonlinear energy bands sum up to a quantized value. This indicates that the geometric contributions from the unique aspects of nonlinear adiabatic following can be important for topological characterization of nonlinear lattice systems. Equally interesting, as nonlinearity strength increases, the individual nonlinear Zak phases, though not quantized, may become closer to a quantized value of either $0$ or $\pi$ than the conventional Zak phases. This further
suggests that nonlinearity can assist in recovering topological effects already destroyed by a chiral-symmetry breaking term. Furthermore, for moderate to strong nonlinearity, clear fingerprints of topological features in the nonlinear system under OBC can be identified.   In-gap localized stationary states (solitons) present nonlinearity induced effective edges inside the bulk. With this understanding, the topological origin of the spatial profiles of such localized solutions can be identified by comparing them with eigenstates of the corresponding linear model under the addition of certain impurity potential. This insight also explains well the recovery of degeneracy of edge solitons localized at opposite ends of the lattice.   %We have provided a unique way to investigate the properties of a nonlinear system by using the topology of the linear system around these effective edges.

\CH{The qualitatively different eigenstates supported by periodic and open boundary conditions also signifies the breaking of bulk-boundary correspondence due to non-linear effects. Fundamentally, this arises because the PBC and OBC solutions no longer span equivalent eigenspaces, with certain solutions i.e. the central spectral loop in Fig 1 existing only at certain Bloch momenta. It remains to be seen how this interplays with an alternative bulk-boundary correspondence breaking mechanism known as the non-Hermitian skin effect~\cite{Yao2018,lee2016anomalous,Lee2019,Leeli2019}, which can also affect the stability of our non-linear eigensolutions had we used another model where $\mathcal{L}$ (Eq.~22) is non-reciprocal in addition to being non-Hermitian.} \GJ{Certainly, our results shall stimulate future work to look into possibly deep connections between our momentum-space results and real-space results, with the long-term hope that new types of bulk-edge correspondence in nonlinear lattice systems can be established.}

As a possible future direction, the interplay between topology and other types of nonlinearity, such as off-diagonal nonlinearity, can be considered as well. We expect that the recovery of topological features in the presence of chiral-symmetry breaking may also be present in such cases, with other potentially more intriguing features yet to be discovered. Moreover, recent years have seen new varieties of exotic topological phases beyond those originally envisioned over two decades ago. These include topological phases in non-equilibrium settings (single-body~\cite{Gong2018PhyRevX,Shen2018,MartinezAlvarez2018,Ghatak:2019zke,Ozdemir2019,Longhi2017,Kawabata2019,Zhou2019NHTopoBands,Kawabata2019NHSym,Li2019NH,Liu2019NH,Lee2019,lee2016anomalous,Yao2018,Leeli2019}, many-body~\cite{yoshida2019non,yoshida2020fate,lee2020ultrafast,zhang2020skin,lee2020many} non-Hermitian and/or periodically driven systems~\cite{Kitagawa2010,Lindner2011,Cayssol2013,Gomez2013FloquetBloch,Grushin2014Floquet,Zhou2014,Perez2015,Asboth2014,Derek2012Floquet,Bomantra2016Floquet,lee2018floquet,Linhu2018Floquet,lee2020quenched}) as well as higher-order topological phases~\cite{Benalcazar2017,BenalcazarHingeState2017,Schindler2018,Song2017,Langbern2017,LiuHOTP2017,Khalaf2018} characterized by the presence of states localized at the boundaries of their boundaries (hinges/corners). Investigating interaction/nonlinear effects in such systems will be timely and fruitful.   
%While we restricted ourselves to the study of a 1D model with diagonal nonlinearity, we expect that more interesting features can be observed in other models, such as those involving off-diagonal nonlinearities or those in 2D or 3D, whose details are left for potential future studies. Another possible future work is the study of the Floquet eigenspectrum of a non-linear 2-mode system under periodic driving, \blue{as well as the study of the non-Hermitian dynamics of the non-linear solutions}.

\begin{acknowledgements}
	{\bf Acknowledgement}: R.W.B is supported by the Australian Research Council Centre of Excellence for Engineered Quantum Systems (EQUS, CE170100009).  J.G. is funded by the Singapore National Research Foundation Grant No. NRF-NRFI2017-04 (WBS No.
R-144-000-378- 281) and by the Singapore Ministry of
Education Academic Research Fund Tier-3 Grant No.
MOE2017-T3-1-001 (WBS. No. R-144-000-425-592).
\end{acknowledgements}

\appendix

\section{Nonlinear adiabatic perturbation theory} \label{app}

We consider a two level Gross-Pitaevskii Hamiltonian
\begin{equation}
\label{eqn:A_GPHamiltonian}
    H(\ket{\Psi}) = h_1 \sigma_x + h_2 \sigma_y + h(\Sigma) \sigma_z 
\end{equation}
where $\Sigma = \abs{\Psi_2}^2 - \abs{\Psi_1}^2$. We start by defining a state $\Phi_a = e^{-if} \Psi_a$ with $a = 1,2$, which corresponds to an element of a projective Hilbert space. The total phase $f$ is taken to capture both dynamical and geometric phases of the state $\ket{\Psi}$. Substituting in Eq.(\ref{eqn:A_GPHamiltonian}) and applying $\sum_{a} \Phi_a^{*} ...$ we obtain (summation of repeated indices being implied)
\begin{equation}
\label{eqn:A_dfoverdt}
    \frac{df}{dt} = i \Phi_a^{*} \frac{d\Phi_a}{dt} - \Phi_a^{*} H_{ab} \Phi_b .
\end{equation}
In this case, the nonlinearity may cause the second term to also contribute to the geometrical phase. We perturbatively expand both $f$ and $\Phi_a$ under an adiabatic parameter $\epsilon$ as
\begin{equation}
\label{eqn:A_AdiabaticExpansion}
    \begin{aligned}
    \frac{df}{dt} &= \alpha_0 + \alpha_1 \epsilon + ... \\
    \Phi_a &= \Phi_a^{(0)} + \epsilon \Phi_a^{(1)} + ...    
    \end{aligned}
\end{equation}
and since the nonlinear Hamiltonian is also state dependent, we will also have
\begin{equation}
\label{eqn:A_PerturbativeExpansionHamiltonian}
    H = H^{(0)} + \epsilon H^{(1)} +  ...
\end{equation}
We now attempt to derive the total phase $f$ acquired by the system in the adiabatic limit for a state initially in a stationary state $\Phi^{(0)}$ such that $H^{(0)} \Phi^{(0)} = E \Phi^{(0)}$, which corresponds to finding $\alpha_0$ and $\alpha_1$ in Eq.(\ref{eqn:A_AdiabaticExpansion}). We obtain
\begin{equation}
\label{eqn:A_GeneralAdiabaticLimit}
    \begin{aligned}
    \alpha_0 &= -E,\\
    \epsilon \alpha_1 &= i \Phi_a^{(0)*} \frac{d\Phi_a^{(0)}}{dt} - \epsilon\Phi_a^{(0)*} H_{ab}^{(1)} \Phi_a^{(0)},
    \end{aligned}
\end{equation}
where the first term in the right hand side of the bottom line corresponds to the conventional Berry connection, and the second term is the geometric contribution coming from the dynamical phase, due to nonlinear dynamics. In our case, we have $H^{(1)} = \left. \frac{dh}{d\Sigma} \right\rvert_{\Sigma = \Sigma^{(0)}} \left. \frac{d\Sigma}{d\epsilon} \right\rvert_{\epsilon = 0} \sigma_z $. Using the normalization condition $\operatorname{Re}(\Phi_a^{(0)*} \Phi_a^{(1)}) = 0$, we have $\left. \frac{d\Sigma}{d\epsilon} \right\rvert_{\epsilon = 0} =  - 4 \operatorname{Re}(\Phi_1^{(0)*} \Phi_1^{(1)})$ so
\begin{equation}
    \label{eqn:A_H1Perturbation}
    \begin{aligned}
    H^{(1)}= -4 \left. \frac{dh}{d\Sigma} \right\rvert_{\Sigma = \Sigma^{(0)}} \operatorname{Re}(\Phi_1^{(0)*} \Phi_1^{(1)}) \sigma_z . 
    \end{aligned}
\end{equation}
The general formula for $\alpha_0$ and $\alpha_1$ given in Eq.(\ref{eqn:A_GeneralAdiabaticLimit}) becomes then
\begin{equation}
		\label{eqn:A_Alpha0and1}
            \begin{aligned}
            \alpha_0 &= -E  \\
             \epsilon \alpha_1 &= i \Phi_a^{(0)*} \frac{d\Phi_a^{(0)}}{dt} - 4 \epsilon \left. \frac{dh}{d\Sigma} \right\rvert_{\Sigma^{(0)}} \Sigma^{(0)} \operatorname{Re}(\Phi_1^{(0)*} \Phi_1^{(1)}) .
            \end{aligned}
\end{equation}
On the other hand, if we consider only $\epsilon^1$ terms in $\frac{df}{dt}\Phi_1$, we have 
\begin{equation}
\label{eqn:A_InjectInGP}
    \begin{split}
    4 \epsilon \left. \frac{dh}{d\Sigma} \right\rvert_{\Sigma^{(0)}} \operatorname{Re}(\Phi_1^{(0)*} \Phi_1^{(1)}) [1 + \Sigma^{(0)}] \Phi_1^{(0)} = \\ -i (\delta_{1a} - \Phi_1^{(0)} \Phi_a^{(0)*})\frac{d\Phi_a^{(0)}}{dt} - \epsilon (E \delta_{1b} - H_{1b}^{(0)}) \Phi_b^{(1)} .
    \end{split}
\end{equation}
For a two-level system, the stationary state $\ket{\Phi_E}$ can be written without loss of generality in the form
\begin{equation}
\label{eqn:A_StationaryPsiE}
    \ket{\Phi_E}=\begin{pmatrix} \cos{\frac{\theta}{2}} \\ \sin{\frac{\theta}{2}}e^{i\phi} \end{pmatrix},
\end{equation}
so that we can simplify Eq.(\ref{eqn:A_InjectInGP}) by taking its real part, and making use again of the normalization condition $\cos{\frac{\theta}{2}} \operatorname{Re}(\Phi_1^{(1)}) + \sin{\frac{\theta}{2}} \operatorname{Re}(e^{-i \phi} \Phi_2^{(1)}) = 0$ to get
\begin{equation}
\label{eqn:A_RealPartInjectInGP}
    \begin{split}
    4 \epsilon \left. \frac{dh}{d\Sigma} \right\rvert_{\Sigma^{(0)}} \cos^2{\frac{\theta}{2}} \operatorname{Re}( \Phi_1^{(1)}) [1 - \cos{\theta}] = \\ 
    i \cos{\frac{\theta}{2}} \Phi_a^{(0)*}\frac{d\Phi_a^{(0)}}{dt} - \epsilon (E - H_{11}^{(0)} + \cot{\frac{\theta}{2}} H_{12}^{(0)} e^{i \phi}) \operatorname{Re}(\Phi_1^{(1)}) .
    \end{split}
\end{equation}
Now we can notice that 
\begin{equation}
		\label{A_eqn:HiddenEigenState}
            \begin{aligned}
            \ket{\Phi^{(0) \perp}} = \begin{pmatrix} \sin{\frac{\theta}{2}} \\ -\cos{\frac{\theta}{2}} e^{i\phi} \end{pmatrix}
            \end{aligned}
\end{equation}
is an (hidden) eigenstate \footnote{$\ket{\Phi^{(0) \perp}}$ is however not a stationary state of the system, as $H^{(0)}$ is state dependent, and $\ket{\Phi^{(0) \perp}}$ is an eigenstate of $H^{(0)}\left(\ket{\Phi^{(0)}}\right)$, but not necessarily of $H^{(0)}\left(\ket{\Phi^{(0) \perp}}\right)$ } of $H^{(0)}$ with eigenvalue $ - E$ \footnote{If we consider the full Bloch space Hamiltonian including the $\frac{g}{2}I_2$ term, $\ket{\Phi^{(0) \perp}}$ then has eigenvalue $ - E + g$, which after calculations, replaces $E$ by $E-\frac{g}{2}$ in Eq.(\ref{eqn:CorrectionNLBP}), effectively cancelling the contribution of the energy shift in the deforming kernel.}, and using this property we obtain after multiplication by $\sin{\frac{\theta}{2}}$
\begin{equation}
    \label{eqn:RealPartofPsi1E>0}
    \epsilon \operatorname{Re}(\Phi_1^{(1)}) = \frac{\cos{\frac{\theta}{2}}}{2E + 2 \left. \frac{dh}{d\Sigma} \right\rvert_{\Sigma^{(0)}} \sin^2{\theta}} i \Phi_a^{(0)*}\frac{d\Phi_a^{(0)}}{dt},
\end{equation}
so subbing in this to Eq.(\ref{eqn:A_Alpha0and1}) gives us the result obtained in Eq.(\ref{eqn:CorrectionNLBP}).

\section{Nonlinear perturbation theory}
\label{app2}

We consider the nonlinear SSH model whose Hamiltonian is given by Eq.(\ref{eqn:BlochSpaceHamiltonian}), and we write it as the sum of a Hamiltonian $H_0$ and a perturbation $V \ll H_0$ 
\begin{equation}
    \label{eqn:B_SplitHamilonianForV}
    H = \underbrace{h_x \sigma_x + h_y \sigma_y + h(\Sigma) \sigma_z}_{H_0} + \underbrace{v \sigma_z}_{V}
\end{equation}
where $h_x = J_1 + J_2 \cos{k}$, $h_y = J_2 \sin{k}$ and $h(\sigma) = \frac{g}{2} \Sigma$ where $\Sigma = \left|\Psi_{2}\right|^2 - \left|\Psi_{1}\right|^2$. Considering a stationary state $\ket{\Psi}$ such that $H\ket{\Psi} = E \ket{\Psi}$, we perturbatively expand both $E$ and $\ket{\Psi}$ under the parameter $v$ as 
\begin{equation}
\label{eqn:B_PerturbativeExpansion}
    \begin{aligned}
    E &= E^{(0)} + v E^{(1)} + ... \\
    \ket{\Psi} &=  \ket{\Psi^{(0)}} + v  \ket{\Psi^{(1)}} + ...    
    \end{aligned}
\end{equation}
Moreover, since $H_0$ is state dependent, we also need to perturbatively expand $H$ as
\begin{equation}
\label{eqn:B_PerturbativeExpansionH}
    \begin{aligned}
    H &= H_0^{(0)} + v H_0^{(^1)} + v \sigma_z + ... \\
    &=  H_0^{(0)} + v \left. \frac{dH_0}{dv}\right\rvert_{v=0} + v \sigma_z + ... \\
    &= H_0^{(0)} + v (1 - 2g\operatorname{Re}(\Psi_1^{(0)*} \Psi_1^{(1)})) \sigma_z + ... 
    \end{aligned}
\end{equation}
Using these perturbative expansions we get by considering only the $v^0$ terms
\begin{equation}
\label{eqn:B_v0Terms}
    H_0^{(0)}\ket{\Psi^{(0)}} = E^{(0)} \ket{\Psi^{(0)}},
\end{equation}
and by considering only the $v^1$ terms
\begin{equation}
    \label{eqn:B_v1Terms}
    (H_0^{(0)} - E^{(0)} )\ket{\Psi^{(1)}} = (E^{(1)} - (1 - 2g\operatorname{Re}(\Psi_1^{(0)*} \Psi_1^{(1)})) \sigma_z  )\ket{\Psi^{(0)}}.
\end{equation}
Eq.(\ref{eqn:B_v0Terms}) is a nonlinear eigenvalue equation that can be solved using the self-consistency equation 
\begin{equation}
\label{eqn:B_SelfConsistency}
    \begin{aligned}
    \left(h_x^2+h_y^2+h(\Sigma^{(0)})^2 \right) \Sigma^2 - h(\Sigma^{(0)})^2 = 0
    \end{aligned}
\end{equation}
which has 4 solutions,
\begin{equation}
\label{eqn:B_SolutionsSelfConsistency}
    \begin{aligned}
    \Sigma^{(0)} = 0 \quad &\textrm{with} \quad E^{(0)} = \pm \sqrt{h_x^2 + h_y^2} \\
    \Sigma^{(0)} = \pm \sqrt{\frac{g^2-4(h_x^2 + h_y^2)}{g^2}} \quad &\textrm{with} \quad E^{(0)} = -\frac{g}{2}.
    \end{aligned}
\end{equation}
The two $\Sigma^{(0)} \neq 0$ solutions are physical only if $g > \sqrt{h_x^2 + h_y^2}$, which we assume to be true as we are interested in the large nonlinearity regime. We can now, without loss of generality, write $\ket{\Psi^{(0)}}$ in the form
\begin{equation}
\label{eqn:B_StationaryPsi}
    \ket{\Psi^{(0)}}=\begin{pmatrix} \cos{\frac{\theta}{2}} \\ \sin{\frac{\theta}{2}}e^{i\phi} \end{pmatrix},
\end{equation}
and plugging this in Eq.(\ref{eqn:B_v1Terms}) after multiplying by $\bra{\Psi^{(0)}}$ gives
\begin{equation}
\label{eqn:B_E1Psi1}
    E^{(1)} = \cos{\theta}(1 - 2g \cos{\frac{\theta}{2}} \operatorname{Re}(\Psi_1^{(1)})).
\end{equation}
Plugging back in Eq.(\ref{eqn:B_v1Terms}) and focusing on the first coefficient, we get (implying summation of repeated indices)
\begin{equation}
\label{eqn:B_PlugInv1Terms}
    \cos{\frac{\theta}{2}} \cos{\theta} (1 - 2g \cos{\frac{\theta}{2}} \operatorname{Re}(\Psi_1^{(1)})) = E^{(0)} \Psi_1^{(1)} - H_{0,1a}^{0} \Psi_a^{(1)},
\end{equation}
and taking the real and making use of the normalization condition $\cos{\frac{\theta}{2}} \operatorname{Re}(\Psi_1^{(1)}) + \sin{\frac{\theta}{2}} \operatorname{Re}(e^{-i \phi} \Psi_2^{(1)}) = 0$ gives us
\begin{equation}
\label{eqn:B_RePsi1}
    \operatorname{Re}(\Psi_1^{(1)}) = \frac{\cos{\frac{\theta}{2}} (1-\cos{\theta} )}{E^{(0)} + g \sin^2{\theta} + \frac{g}{2} \cos{\theta} + \cot{\frac{\theta}{2}} \sqrt{h_x^2 + h_y^2} }
\end{equation}
so plugging in Eq.(\ref{eqn:B_E1Psi1}), we have the first order correction to the energy
\begin{equation}
\label{eqn:B_E1}
    E^{(1)} = \cos{\theta} \left( 1 - \frac{g \sin^2{\theta}}{E^{(0)} + g \sin^2{\theta} + \frac{g}{2} \cos{\theta} + \cot{\frac{\theta}{2}} \sqrt{h_x^2 + h_y^2} } \right).
\end{equation}
Now for $\ket{\Psi^{(1)}}$, we have 
\begin{equation}
\label{eqn:B_Psi}
    \begin{aligned}
    \ket{\Psi} &=  \ket{\Psi^{(0)}} + v  \ket{\Psi^{(1)}} \\
    \begin{pmatrix} \cos{\frac{\theta'}{2}} \\ \sin{\frac{\theta'}{2}}e^{i\phi} \end{pmatrix} &= \begin{pmatrix} \cos{\frac{\theta}{2}} \\ \sin{\frac{\theta}{2}}e^{i\phi} \end{pmatrix} + v \begin{pmatrix} \Psi_1^{(1)} \\ \Psi_2^{(1)} \end{pmatrix}
    \end{aligned}
\end{equation}
so $\cos{\frac{\theta'}{2}} = \cos{\frac{\theta}{2}} + v \Psi_1^{(1)}$ tells us that $\Psi_1^{(1)}$ is real, i.e., $\Psi_1^{(1)} = \operatorname{Re}(\Psi_1^{(1)})$. We then consider once again Eq.~(\ref{eqn:B_PlugInv1Terms}), this time taking the imaginary part, to show that $\operatorname{Im}(e^{-i\phi} \Psi_2^{(1)}) = 0$, so $\Psi_2^{(1)}$ can be written $\Psi_2^{(1)} = \sin{\frac{\theta_1}{2}} e^{i\phi} $. Using the normalization condition, we get $\sin{\frac{\theta_1}{2}} = -\cot{\frac{\theta}{2}} \operatorname{Re}(\Psi_1^{(1)})$. This gives us the first order correction to the stationary state
\begin{equation}
\label{eqn:B_Psi1}
     \begin{pmatrix} \Psi_1^{(1)} \\ \Psi_2^{(1)} \end{pmatrix} = \begin{pmatrix} \frac{\cos{\frac{\theta}{2}} (1-\cos{\theta} )}{E^{(0)} + g \sin^2{\theta} + \frac{g}{2} \cos{\theta} + \cot{\frac{\theta}{2}} \sqrt{h_x^2 + h_y^2} } \\ -\frac{\cot{\frac{\theta}{2}}\cos{\frac{\theta}{2}} (1-\cos{\theta} )}{E^{(0)} + g \sin^2{\theta} + \frac{g}{2} \cos{\theta} + \cot{\frac{\theta}{2}} \sqrt{h_x^2 + h_y^2} } e^{i\phi} \end{pmatrix}.
\end{equation}
Now that $E^{(1)}$ and $\ket{\Psi^{(1)}}$ have been determined, we can compute the nonlinear Zak phase. In order to make it analytically calculable, we assume $J_1 = 0$ and $J_2 \ll g$, doing all the perturbative expansions up to $\mathcal{O}\left( \frac{v J_2^2}{g^3} \right)$. This way it is possible to determine the  new states $\ket{\Psi}$ and the new energy $E$, along with the deforming kernel $\mathcal{K}$. After some analysis, we get for the different energy bands $E_1<E_2<E_3<E_4$,
\begin{equation}
    \label{eqn:B_AllKs_Connections}
    \begin{aligned}
        \mathcal{K}_1 &= 1 + 2 \left( \frac{J_2}{g} \right)^2 + \mathcal{O}\left(\frac{v J_2^2}{g^3}
        \right) \\
        \mathcal{K}_2 &= -1 -4\frac{v}{g} -2 \left( \frac{J_2}{g} \right)^2 + \mathcal{O}\left(\frac{v J_2^2}{g^3} \right) \\
        \mathcal{K}_3 &= 1 +2 \frac{v}{g} + 8 \frac{v J_2}{g^2}  + \mathcal{O}\left(\frac{v J_2^2}{g^3} \right) \\
        \mathcal{K}_4 &= 1 +2 \frac{v}{g} - 8 \frac{v J_2}{g^2} + \mathcal{O}\left(\frac{v J_2^2}{g^3} \right), \\
        \textnormal{and} \\
        i \bra{\Psi_{E_1}(k)}\nabla_{k}\ket{\Psi_{E_1}(k)} &= -(1 - \left( \frac{J_2}{g} \right)^2) + \mathcal{O}\left(\frac{v J_2^2}{g^3}
        \right) \\
        i \bra{\Psi_{E_2}(k)}\nabla_{k}\ket{\Psi_{E_2}(k)} &= - \left( \frac{J_2}{g} \right)^2 + \mathcal{O}\left(\frac{v J_2^2}{g^3} \right) \\
        i \bra{\Psi_{E_3}(k)}\nabla_{k}\ket{\Psi_{E_3}(k)} &= -\frac{1}{2} (1 - 2 \frac{v}{g} - 4 \frac{v J_2}{g^2} ) + \mathcal{O}\left(\frac{v J_2^2}{g^3} \right) \\
        i \bra{\Psi_{E_4}(k)}\nabla_{k}\ket{\Psi_{E_4}(k)} &= -\frac{1}{2} (1 - 2 \frac{v}{g} + 4 \frac{v J_2}{g^2} ) + \mathcal{O}\left(\frac{v J_2^2}{g^3} \right),
    \end{aligned}
\end{equation}
which gives the nonlinear Zak phases presented in Eq.~(\ref{eqn:PerturbativeZakPhase}).

%\twocolumngrid
\bibliographystyle{apsrev4-2}
%\begin{thebibliography}{103}
%\bibliography{Bibliography}

\end{document}